\documentclass[11pt]{article}
\usepackage{fancyhdr,psfig,lastpage}
\usepackage{amsmath}
\usepackage{epsfig,latexsym,cite}
\begin{document}

\vspace*{0.8cm}
\begin{center}
{\Large \bf Uphill Motion of Active Brownian Particles \\ 
in Piecewise Linear
  Potentials}\\[10mm]

{\large Frank Schweitzer$^{1\star}$, Benno Tilch$^{2}$,  Werner Ebeling$^{1}$}\\[4mm]

\begin{quote}
\begin{itemize}
\item[$^{1}$]{\it Institute of Physics, Humboldt University Berlin,
  Invalidenstra{\ss}e 110, D-10115 Berlin, Germany}
\item[$^{2}$]{\it II.  Institute of
  Theoretical Physics, University of Stuttgart, Pfaffenwaldring 57/III,
  D-70550 Stuttgart, Germany} 
\item[$\star$] Present address: \emph{
GMD Institute for Autonomous intelligent
      Systems, Schloss Birlinghoven, 53754 Sankt Augustin, Germany,
      e-mail: schweitzer@gmd.de}
\end{itemize}
\end{quote}
\end{center}

\begin{abstract}
We consider Brownian particles with the ability to take up energy from
  the environment, to store it in an internal depot, and to convert
  internal energy into kinetic energy of motion. Provided a supercritical
  supply of energy, these particles are able to move in a ``high
  velocity'' or active mode, which allows them to move also against the
  gradient of an external potential.  We investigate the critical
  energetic conditions of this self-driven motion
for the case of a linear potential and a ratchet
  potential.  In the latter case, we are able to find two different
  critical conversion rates for the internal energy, which describe the
  onset of a directed net current into the two different directions. The
  results of computer simulations are confirmed by analytical expressions
  for the critical parameters and the average velocity of the net
  current. Further, we investigate the influence of the asymmetry
  of the ratchet potential on the net current and estimate a
  critical value for the asymmetry in order to obtain a positive or
  negative net current.
\end{abstract} 

05.45.-a Nonlinear dynamics and nonlinear dynamical systems \\
05.60.-k Transport processes \\
82.20.Mj Nonequilibrium kinetics \\
87.10.+e General theory and mathematical aspects

\newcommand{\bbox}[1]{\mbox{\boldmath $#1$}}
\newcommand{\mean}[1]{\left\langle #1 \right\rangle}
\newcommand{\abs}[1]{\left| #1 \right|}
\newcommand{\la}{\langle}
\newcommand{\ra}{\rangle}
\newcommand{\RA}{\Rightarrow}
\newcommand{\tet}{\vartheta}
\newcommand{\eps}{\varepsilon}
\renewcommand{\phi}{\varphi}
\newcommand{\bib}[2]{\bibitem{#1}{\rm #2.}}
\newcommand{\non}{\nonumber \\}
\newcommand{\no}{\nonumber}
\newcommand{\sect}[1]{Sect. \ref{#1}}
\newcommand{\eqn}[1]{Eq. (\ref{#1})}
\newcommand{\eqs}[2]{Eqs. (\ref{#1}), (\ref{#2})}
\newcommand{\pic}[1]{Fig. \ref{#1}}
\newcommand{\vol}[1]{{\bf #1}}
\newcommand{\et}{{\it et al.}}
\newcommand{\fn}[1]{\footnote{ #1}}
\newcommand{\name}[1]{{\rm #1}}
\newcommand{\D}{\displaystyle}
\newcommand{\T}{\textstyle}
\newcommand{\SC}{\scriptstyle}
\newcommand{\SSC}{\scriptscriptstyle}

\renewcommand{\textfraction}{0.02}
\renewcommand{\topfraction}{0.98}
\renewcommand{\bottomfraction}{0.98}
\renewcommand{\floatpagefraction}{0.98}


\section{Introduction}  
\label{1}

The motion of a ``simple'' Brownian particle is due to fluctuations of
the surrounding medium, i.e. the result of random impacts of the
molecules or atoms of the liquid or gas, the particle is immersed in.
This type of motion would be rather considered as \emph{passive motion},
simply because the Brownian particle does not play an active part in this
motion. On the other hand, already %
in physico-chemical systems a \emph{self-driven motion} of particles can
be found \cite{Mikhailov-Meinkoehn-97}. For instance, small solid
particles floating on a liquid surface may produce a chemical substance
at a spatially inhomogeneous rate, which locally changes the surface
tension. This way the substance induces a net capillary force acting on
the particle, which results in the particle's motion.

Recent investigations on \emph{interacting} self-driven particles show a
broad variety of interesting phenomena, such as phase transitions and the
emergence of self-ordered motion
\cite{vicsek-et-95,albano-96,helbing-vicsek-99}. Here, the focus is on
collective effects rather than on the origin of the particle's 
velocity; i.e. it is usually postulated that the particles move with a
certain non-zero velocity. 

In this paper, we focus on the \emph{energetic} aspects of self-driven
motion in order to derive conditions for an \emph{active mode} of motion.
\emph{Active motion}, as the name suggests, occurs under energy
consumption, it is related to processes of energy storage and conversion
into kinetic energy.  For instance on the biological level, cells or
simple microorganisms are capable of \emph{active}, self-driven motion,
which in many cases has been successfully described by stochastic
equations
\cite{tranq-lauffenb-87,othmer-et-88,alt-hoffm-90,dickinson-tranq-93,schienb-gruler-93,vicsek-et-95}.

In order to describe both the \emph{random} aspects and the
\emph{energetic} aspects of active motion, 
we have introduced a model of {\em active
  Brownian particles}
\cite{fs-lsg-94,steuern-et-94,lsg-mieth-rose-malch-95,
  fs-agent-97,fs-eb-tilch-98-let,eb-fs-tilch-98}. 
These are Brownian particles with the 
ability to take up energy from the environment, to store it in an
internal depot and to convert internal energy to perform different
activities, such as metabolism, motion, change of the environment, or
signal-response behavior. Since the focus in this paper is on the
energetic aspects of active Brownian particles in a
specific potential, possible changes of the environment are neglected
here.

A very simple mechanism to take-up the additional energy required for
active motion, is the pumping of energy by \emph{space-dependent
  friction} \cite{steuern-et-94}. In this case, the friction coefficient
$\gamma_{0}$ becomes a space-dependent function, $\gamma(\bbox{r})$,
which in a certain spatial range can be also negative. Inside this area
the Brownian particle, instead of loosing energy because of dissipative
processes, is pumped with energy, which in turn increases its velocity.
Provided a supercritical supply of energy, the particle should be able to
move with a velocity larger than the average thermal velocity.

While such an approach will be able to model the spatially inhomogeneous
supply of energy, it has the drawback not to consider processes of
storage and conversion of energy. In fact, with only a space-dependent
friction, the Brownian particle is instantaneously accelerated or slowed
down, whereas e.g. biological systems have the capability to stretch their
supply of energy over a certain time interval.

In order to develop a more realistic model of active motion, we have
considered an \emph{internal energy depot} for the Brownian particles
\cite{fs-eb-tilch-98-let,eb-fs-tilch-98}, which allows to store the
taken-up energy in the internal depot, from where it can be converted
e.g. into kinetic energy, namely for the acceleration of motion.
Additionally, the internal dissipation of energy, due to storage and
conversion (or metabolism in a biological context) can be considered.

With these extensions, the Brownian particle becomes in fact a
\emph{Brownian motor} \cite{magnasco-94,astumian-bier-94,juelicher-prost-95}, which is fueled somewhere and then uses
the stored energy with a certain efficiency \cite{eb-fs-tilch-98} to move
forward, also against external forces. Provided a supercritical supply of
energy, we have found that the motion of active Brownian particles in the
two-dimensional space can become rather complex
\cite{fs-eb-tilch-98-let}.

In this paper we investigate the one-dimensional motion of an ensemble
of Brownian particles with internal energy depot in piecewise linear
potentials. One particular example is the ratchet potential,
i.e.  a periodic potential which lacks the reflection symmetry. Ratchet
systems recently attracted much interest with respect to transport
phenomena on the microscale, since they provide a mechanism to transfer
the undirected motion of Brownian particles into a directed motion.
Hence, the term {\em Brownian rectifiers} \cite{haenggi-bart-96} has been
established.  In order to reveal the microscopic mechanisms resulting in
directed movement, different physical
ratchet models have been proposed \cite{
maddox-nat-94b}, such as {\em forced
  thermal ratchets} \cite{magnasco-93
  }, or {\em stochastic ratchets}
\cite{luczka-bart-haenggi-95,millonas-dykman-94}, or {\em fluctuating
  ratchets}
\cite{rousselet-nat-94,zuercher-doering-93
  }.

The model discussed in this paper, aims to add a new perspective to this
problem which is based on the idea of internal energy storage. In
\sect{2}, we discuss the basic features of our model without specifying
an external potential, and derive the equations of motion both for the
general case and the overdamped limit. In particular, we point to the
possible existence of a ``high velocity'' or \emph{active mode} of motion
for the Brownian particle, in addition to the usual passive mode of
motion.  In \sect{3}, the necessary conditions for such an active mode of
motion are investigated for the case of a linear potential. For the
overdamped limit, we derive critical parameters of energy conversion,
which allows the particle to move also \emph{against} the direction of
the external force, i.e. to move ``uphill'' the potential gradient. The
results are applied to the motion of Brownian particles in a ratchet
potential in \sect{4}.  After investigating the deterministic motion of a
single particle in \sect{4.1}, we discuss the energetic conditions for
the establishment of a net current for the determinsitic motion of an
ensemble of particles in \sect{4.2}. We show that, dependent on the
conversion of internal into kinetic energy, a net current into different
directions can be established. The critical parameters for the energy
conversion and the resulting velocity of the net current, which are found
by means of computer simulations, are compared with the analytical
results obtained in \sect{2}. We also investigate the influence of the
asymmetry of the ratchet potential on the establishment of the net
current.

\section{Equations of Motion of Pumped Brownian Dynamics}  
\label{2}

The motion of simple Brownian particles in a space-dependent potential,
$U(\bbox{r})$ can be described by the Langevin equation:
\begin{equation}
  \label{langev-or}
  \dot{\bbox{r}}=\bbox{v}\,;\,\, 
m\,\dot{\bbox{v}}=-\gamma_{0} \bbox{v} - \nabla U(\bbox{r}) 
+ {\cal F}(t)
\end{equation}
where $\gamma_{0}$ is the friction coefficient of the particle at position
$\bbox{r}$, moving with velocity $\bbox{v}$. ${\cal F}(t)$ is a
stochastic force with strength $D$ and a $\delta$-correlated time
dependence
\begin{equation}
  \label{stoch}
\mean{{\cal F}(t)}=0 \,;\,\,
\mean{{\cal F}(t){\cal F}(t')}=2D \,\delta(t-t')
\end{equation}
Using the fluctuation-dissipation theorem, we
assume that the loss of energy resulting from friction, and the gain of
energy resulting from the stochastic force, are compensated in the
average, and $D$ can be expressed as: 
\begin{equation}
  \label{fluct-diss}
D=k_{B}T\gamma_{0}  
\end{equation}
where $T$ is the temperature and $k_{B}$ is the \name{Boltzmann}
constant.

In addition to the dynamics described above, the Brownian particles
considered here are pumped with energy from the environment, which they
can store in an internal depot. Further, internal energy can be converted
into kinetic energy. Considering also internal dissipation, the resulting
balance equation for the internal energy depot, $e$, of a pumped Brownian
particle is given by:
\begin{equation}
\frac{d}{dt} e(t) = q(\bbox{r}) - c\;e(t) - d(\bbox{v})\;e(t)
\label{dep0}
\end{equation}
$q(\bbox{r})$ is the space-dependent pump rate of energy and $c$ describes
the internal dissipation assumed to be proportional to the depot energy.
$d(\bbox{v})$ is the rate of conversion of internal into kinetic energy
which should be a function of the actual velocity of the particle. A
simple ansatz for $d(\bbox{v})$ reads:
\begin{equation}
d(\bbox{v}) = d_2 v^2
\label{dv}
\end{equation}
where $d_2>0$ is the conversion rate of internal into kinetic energy.
The energy conversion results in an additional acceleration of the
Brownian particle in the direction of movement, expressed by the vector
$\bbox{e}_v=\bbox{v}/v$. Hence, the equation of motion for the pumped
Brownian particles has to consider an additional driving force,
$d_{2}e(t)\bbox{v}$. In \cite{fs-eb-tilch-98-let,eb-fs-tilch-98}, we have
postulated a stochastic equation, which is consistent with the
\name{Langevin} \eqn{langev-or}:
\begin{equation}
  \label{langev-pump}
  \dot{\bbox{r}}=\bbox{v}\,;\,\, 
m\dot{\bbox{v}}+\gamma_{0} \bbox{v} +\nabla U(\bbox{r})
=d_{2}e(t)\bbox{v} + {\cal F}(t)
\end{equation}
If we restrict ourselves to the one-dimensional case, i.e. the space
coordinate is given by $\bbox{x}$ and further assume $m=1$ for the mass
and $q(\bbox{r})=q_0=const.$ for influx of energy, the dynamics for the
pumped Brownian motion, eqs.  (\ref{langev-pump}), (\ref{dep0}) can be
specified as follows:
\begin{eqnarray}
\label{model}
\dot{\bbox{x}} & = & \bbox{v} \nonumber \\
\dot{\bbox{v}} & = & - \Big(\gamma_{0} -d_2 e(t)\Big)\bbox{v} -
\frac{\partial U(x)}{\partial x} + \sqrt{2\,D} \,\bbox{\xi}(t) \\ 
\dot{e} & = & q_0 - c e - d_2 v^2 e \nonumber
\end{eqnarray}
where $\bbox{\xi}$ is a stochastic force with white-noise
fluctuations:
\begin{equation}
\mean{\bbox{\xi}(t)\bbox{\xi}(t')} = \delta(t-t')
\label{noise}
\end{equation}
In order to find a solution for the coupled Eqs. (\ref{model}), let us
now consider a relaxation of the dynamics on different time scales. If we
assume that the velocity $\bbox{v}(t)$ is changing much faster than the
space coordinate $\bbox{x}(t)$, for $\bbox{v}(t)$, \eqn{model}, a formal
solution can be given: 
\begin{eqnarray}
  \bbox{v}(t) & = & \bbox{v}(0)\exp \left\{-\gamma_{0} t 
 + d_2 \int_0^t e(t')
  dt'\right\}\, \nonumber\\ & &+ \, \exp \left\{-\gamma_{0}t+d_2 \int_0^t 
e(t') dt'\right\} \nonumber \\  
& & \times  \int_0^t \exp \left\{\gamma_{0} t' - d_2 \int_0^{t'} e(t'')
  dt''\right\} \nonumber \\  && \times
  \left[-\bbox{\nabla} U + \sqrt{2k_{B}T \gamma_{0}}\,\bbox{\xi}(t')\right] dt'
\label{v-adiab}
\end{eqnarray}
This solution however depends on the integrals over $e(t)$, reflecting
the influence of the energy depot on the velocity. If we further assume a
fast relaxation of the depot, $e(t)$, compared to the relaxation of the
velocity $\bbox{v}(t)$, the corresponding equation of \eqn{model} can be
solved, and we obtain with the initial condition $e(0)=0$ the following
quasistationary value for the energy depot:
\begin{equation}
e_{0} = \frac{q_0}{c+d_2\bbox{v}^2}
\label{e0}
\end{equation}
which yields a possible maximum value of $e_{0}^{max}=q_{0}/c$.  

The overdamped limit is obtained by considering a fast relaxation of the
velocities, in which case the set of \eqn{model} can be further reduced
to:
\begin{equation}
\label{x-overd}
\bbox{v}(t) = - \frac{1}{\gamma_{0} 
- d_2 e_{0}}\;\frac{\partial U}{\partial \bbox{x}} 
+ \frac{\sqrt{2k_{B}T\gamma_{0}}}{\gamma_{0}-d_2e_{0}} \; \bbox{\xi}(t)
\end{equation}
We note that, due to the dependence of $e_{0}$ on
$\bbox{v}^{2}=\dot{\bbox{x}}^{2}$, \eqn{x-overd} is coupled to \eqn{e0}.
Thus, the overdamped \eqn{x-overd} could be also written in the form:
\begin{equation}
\label{x-overd-2}
\left(\gamma_{0} 
- d_2 \frac{q_{0}}{c+d_{2}\,\dot{x}^{2}} \right) \;\dot{\bbox{x}}
= - \frac{\partial U}{\partial \bbox{x}} 
+ \sqrt{2k_{B}T\gamma_{0}} \; \bbox{\xi}(t)
\end{equation}
\eqn{x-overd-2} indicates a cubic equation for the velocities in the
overdamped limit, i.e. the possible existence of non-trivial solutions
for the stationary velocity. For the further discussion, we neglect the
stochastic term in \eqn{x-overd-2} and denote the stationary values of
$\bbox{v}(t)$ by $\bbox{v}_{0}(\bbox{x})$. Further, the force resulting
from the gradient of the potential, \mbox{$\bbox{F}(\bbox{x})=
  -\bbox{\nabla} U$}, is introduced. Then, \eqn{x-overd-2} can be
rewritten as:
\begin{equation}
\Big[d_{2}\gamma_{0}\, v_0^2 - d_2 \bbox{F} \bbox{v}_{0} -(q_{0} d_2
 -c\gamma_{0})\Big]\,  \bbox{v}_0 = c \bbox{F}. 
\label{v0-cubic}
\end{equation}
Depending on the value of $\bbox{F}$ and in particular on the sign of the
term $(q_{0} d_2 -c\gamma)$, \eqn{v0-cubic} has either one or three
real solutions for the statio\-nary velocity, $\bbox{v}_{0}$. The always
existing solution expresses a direct response to the force in the form:
\begin{equation}
  \label{v-norm}
  \bbox{v}_{0}(\bbox{x}) \sim \bbox{F}(\bbox{x}) 
\end{equation}
This solution results from the analytic continuation of Stokes' law,
$\bbox{v}_{0}= \bbox{F}/\gamma_{0}$, which is valid for $d_{2}=0$.
We will denote this solution as the ``normal response'' mode of
motion, since the velocity $\bbox{v}$ has the same direction as the force
$\bbox{F}$ resulting from the external potential $U(\bbox{x})$. 

As long as the supply of the energy depot is small, we will also name the
normal mode as the \emph{passive mode}, because the particle is simply
driven by the external force. 
More interesting is the case of three stationary velocities,
$\bbox{v}_{0}$, which significantly depends on the (supercritical)
influence of the energy depot. In this case which will be discussed in
detail in the following section, the particle will be able to move in a
``high velocity'' or \emph{active mode} of motion.
For the one-dimensional motion, in the active mode only two different
directions are possible, i.e. a motion into or \emph{against} the
direction of the force $\bbox{F}$.  
 
But already in the
two-dimensional case there are infinitely different possibilities. This
conclusion is of importance when discussing stochastic influences as will
be done in a forthcoming paper \cite{ebeling-et-99}. In the
one-dimensional case, the influence of noise is rather weak, because of
the limited number of possible directions, but in the two- and
three-dimensional case, the active motion of the particles is very
sensitive to stochastic influences, which may determine the direction of
motion in the active mode.

\section{Deterministic Motion in a Linear Potential} 
\label{3}
\subsection{Stationary Solutions}
\label{3.1}
In the following, we restrict the discussion to the one-dimensional,
deterministic motion of the particle, corresponding to $D=0$ in
\eqn{model}. Further, we may assume that the force resulting from the
gradient of the potential is constant.
Then, we have the two coupled equations for
$\bbox{v}(t)$ and $e(t)$:
\begin{eqnarray}
\label{model2}
\dot{\bbox{v}} & = & - \Big(\gamma_{0} -d_2 e(t)\Big)\bbox{v} + \bbox{F} \\
\dot{e} & = & q_0 - c e - d_2 \bbox{v}^2 e \nonumber
\end{eqnarray}
The stationary solutions of \eqn{model2} are obtained from
$\dot{\bbox{v}}=0$ and $\dot{e}=0$:
\begin{equation}
\bbox{v}_0  =  \frac{\bbox{F}}{\gamma_{0} -d_2 e_0} \quad \mbox{;}\quad
e_0  =  \frac{q_0}{c + d_2 \bbox{v}_0^{2}}
\label{v0-e0}
\end{equation}
which lead to the known cubic polynom for the amount of the constant
velocity $v_{0}$, \eqn{v0-cubic}: 
\begin{equation}
d_{2}\gamma_{0} \bbox{v}_0^3 - d_2 \bbox{F} v_0^2 
-(q_0 d_2 -c\gamma_{0}) \bbox{v}_0 -c \bbox{F} =0.
\label{v0-cubic-2}
\end{equation}
Here $\bbox{v}_{0}^{n+1}$ is defined as a  vector
$\abs{v_{0}}^{n}\bbox{v}_{0}$. 

For a first insight into the possible solutions of \eqn{v0-cubic-2}, we
consider the force free motion, $\bbox{F}=0$. Then \eqn{v0-cubic-2} has
either one or three stationary solutions which read:
\begin{equation}
  \label{3stat-F0}
v_{0}^{(1)}=0\;\;,\quad
v_{0}^{(2,3)} = \pm
\sqrt{\frac{q_0}{\gamma_{0}}-\frac{c}{d_{2}}} 
\end{equation}
The corresponding bifurcation diagram is shown in \pic{bifurc}a which
displays a typical fork bifurcation. The bifurcation point is given by: 
\begin{equation}
  \label{bif-point}
  d_{2}^{bif}=\frac{c\,\gamma_{0}}{q_{0}}
\end{equation}
Above a critical supply of energy, which is expressed in terms of the
conversion parameter $d_{2}$, we find the occurence of two new solutions
for the stationary velocity, corresponding to the active modes of motion.
For $\bbox{F}=0$, $v_{0}^{(2)}$ and $v_{0}^{(3)}$ both have the same
amount but different directions. 
Below the critical threshold the passive mode of motion, $v_{0}^{(1)}=0$,
is the only stable solution. At the bifurcation point, a change of the
stability occurs, consequently above the critical value of $d_{2}$ the
active modes of motion are assumed stable.
\begin{figure}[htbp]
 \centerline{\psfig{figure=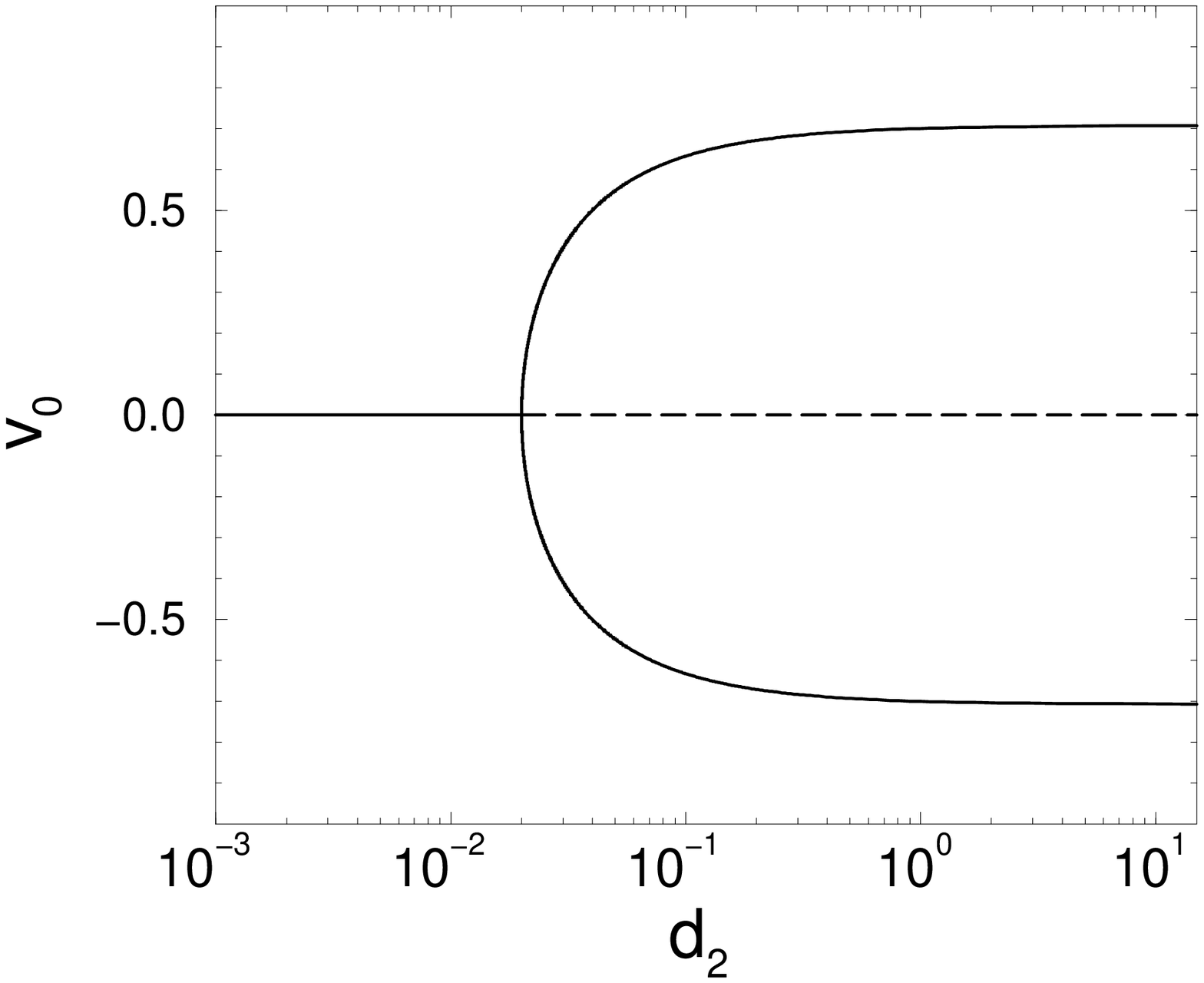,width=7.5cm}}
 \centerline{\psfig{figure=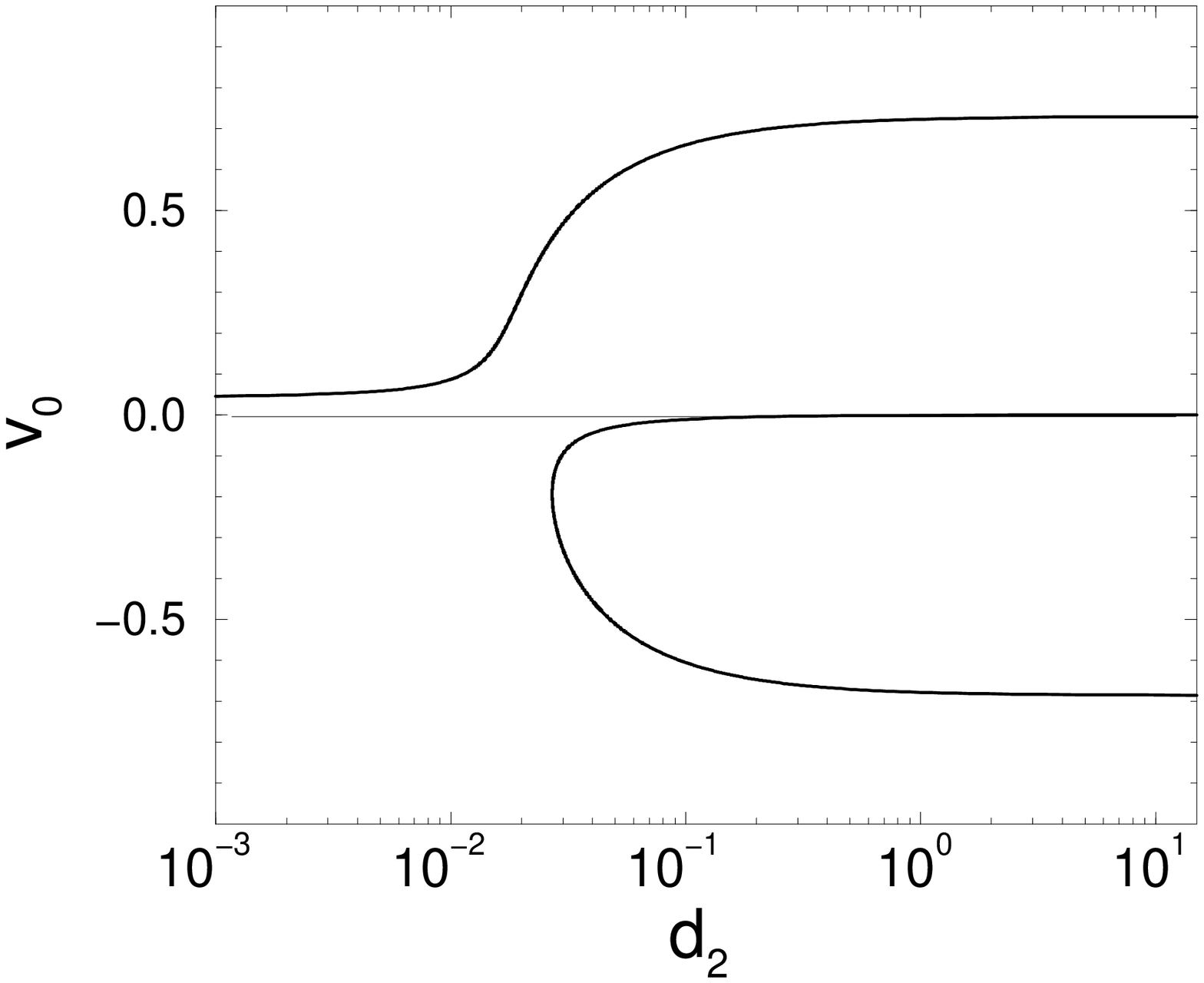,width=7.5cm}}
\caption[fig2]{
  Stationary velocities $v_{0}$, \eqn{v0-cubic-2}, vs. conversion rate
  $d_{2}$: (a: top) for $\bbox{F}=0$, (b: bottom) for $\bbox{F}=+7/8$.
  Above a critical value of $d_{2}$, a negative stationary velocity
  indicates the posssibility to move against the direction of the force.
  Parameters: $q_{0}=10$, $\gamma_{0}=20$, $c=0.01$.
  \label{bifurc}}
\end{figure}

For the case $\bbox{F}=const. \neq 0$, the analysis of \eqn{v0-cubic-2}
leads to a different bifurcation diagram shown in \pic{bifurc}b. Above
a critical supply of energy, we find the appearance of two high velocity
or active modes of motion.  One of these active modes has the same
direction as the driving force, thus it can be understood as the
continuation of the normal solution.  As \pic{bifurc}b shows, the former
passive normal mode, which holds for subcritical energetic conditions, is
transformed into an active normal mode, where the particle moves into the
same direction, but with a much higher velocity.  \emph{Additionally}, in
the active mode a new high-velocity motion \emph{against} the direction
of the force $\bbox{F}$ becomes possible. While the first active mode
would be rather considered as a \emph{normal response} to the force
$\bbox{F}$, the second active mode appears as unnormal (or non-trivial)
response, which corresponds to an ``uphill'' motion (cf.  \pic{flanke1}).

It is obvious that the particle's motion ``downhill'' is stable, but the
same does not necessarily apply for the possible solution of an
``uphill'' motion. Thus, in addition to \eqn{v0-cubic-2} which provides
the \emph{values} of the stationary solutions, we need a second condition
which guarantees the \emph{stability} of these solutions. Before this is
carried out, we want to derive a handy expression for the stationary
velocities in the case $\bbox{F}\neq 0$. With the assumption that the
term $c\bbox{F}$ is small, the stationary solutions of \eqn{v0-cubic-2}
can be given as:
\begin{equation}
  \label{3stat-c0}
v_{0}^{(1)}=0\;\;,\quad
v_{0}^{(2,3)} = \frac{F}{2\gamma_{0}} \pm
\sqrt{\frac{F^2}{4\gamma_{0}^2}
  +\left(\frac{q_0}{\gamma_{0}}-\frac{c}{d_{2}}\right)} 
\end{equation}
We note that \eqn{3stat-c0} is a sufficient approximation for the
stationary velocities, especially in a certain distance from the
bifurcation point, thus it will be also used in \sect{4.1}.
\begin{figure}[htbp]
\vspace{-0.2cm}
 \centerline{\psfig{figure=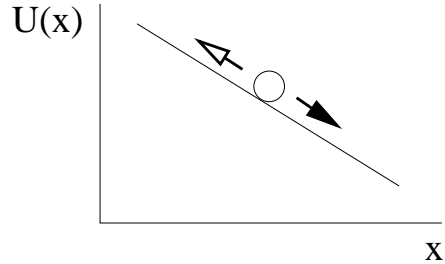,height=3.5cm}}
\caption[fig1]{
 Sketch of the one-dimensional deterministic motion of the particle in
 the presence of a constant force $\bbox{F}=-\bbox{\nabla}U(x)=const.$.
 Provided a supercritical amount of energy from the depot, the particle
 might be able to move ``uphill'', i.e. against the direction of the
 force.
 \label{flanke1}}
\end{figure}

\subsection{Stability Analysis}
\label{3.2}
For the stability analysis, we consider small fluctuations around the
stationary values, $v_{0}$ and $e_{0}$:
\begin{equation}
  \label{fluct}
v=v_{0}+\delta v\;;\;\; e=e_{0}+\delta e \;;\;\;
\abs{\frac{\delta v}{v_{0}}} \sim \abs{\frac{\delta e}{e_{0}}} \ll 1  
\end{equation}
Inserting \eqn{fluct} into \eqn{model2}, we find after linearization:
\begin{equation}
  \label{lin-stab}
  \begin{array}{rllll}
  \dot{\delta v} &=& \delta v \;\Big(-\gamma_{0} + d_{2} e_{0}\Big)
&+& \delta e \;\Big(d_{2} v_{0} \Big) \\
 \dot{\delta e} &=& \delta v \;\Big(-2 d_{2} e_{0} v_{0}\Big)
&+& \delta e \;\Big(-c - d_{2} v_{0}^{2} \Big) 
  \end{array}
\end{equation}
With the ansatz:
\begin{equation}
  \label{e-ans}
  \delta v \sim \sim \exp{\{\lambda t\}}\;;\;\;
\delta e \sim \exp{\{\lambda t\}}
\end{equation}
we find from \eqn{lin-stab} the following relation for $\lambda$:
\begin{eqnarray}
  \label{lamb-12}
  \lambda^{(1,2)} &=& -\frac{1}{2} \Big(\gamma_{0}+ c+
  d_{2}v_{0}^{2}-d_{2}e_{0}\Big) \non 
&& \pm \sqrt{\frac{1}{4}\Big(\gamma_{0}+ c+
  d_{2}v_{0}^{2}-d_{2}e_{0}\Big)^{2} - \cdots } \non
&&\quad \overline{\cdots-c(\gamma_{0}-d_{2}e_{0})^{\rule{0cm}{8pt}}
-d_{2}v_{0}^{2}(\gamma_{0} +d_{2}e_{0})} 
\end{eqnarray}
In general, we need to discuss \eqn{lamb-12} for the three
possible solutions $v_{0}$ which result from \eqn{v0-cubic-2}.  Dependent
on whether the $\lambda$ for each solution have real or complex positive
or negative values, we are able to classify the types of the possible
stationary solutions in this case.  The results are summarized in Table
\ref{tab-bifurc}. The phase plots shown in \pic{phase}a,b present more
details. Further, \pic{lambda12} shows the real part $\Re (\lambda)$ of
\eqn{lamb-12} for the active mode corresponding to the ``uphill'' motion
of the particle, which is the most interesting one.

\begin{table}[htbp]
  \begin{center}
    \begin{tabular}[t]{rllcp{4.cm}} \hline \hline
0 & $ < d_{2} < $ & 0.027   &  : & 1 stable node \\
0.027 & $< d_{2}  < $&  0.046 &: & 1 stable node, 1 instable node, 1
saddle point\\  
0.046 & $< d_{2} < $& 2.720 & :& 1 stable focal point,  1 instable focal
point, 1 saddle point\\  
2.720 & $< d_{2}  $ &    &  : & 2 stable focal points, 1 saddle point \\
\hline \hline 
    \end{tabular}\bigskip
    \caption{Results for the stability analysis, \eqn{lamb-12}.
    \label{tab-bifurc}}
  \end{center}
\end{table}

\begin{figure}[htbp]
\centerline{\psfig{figure=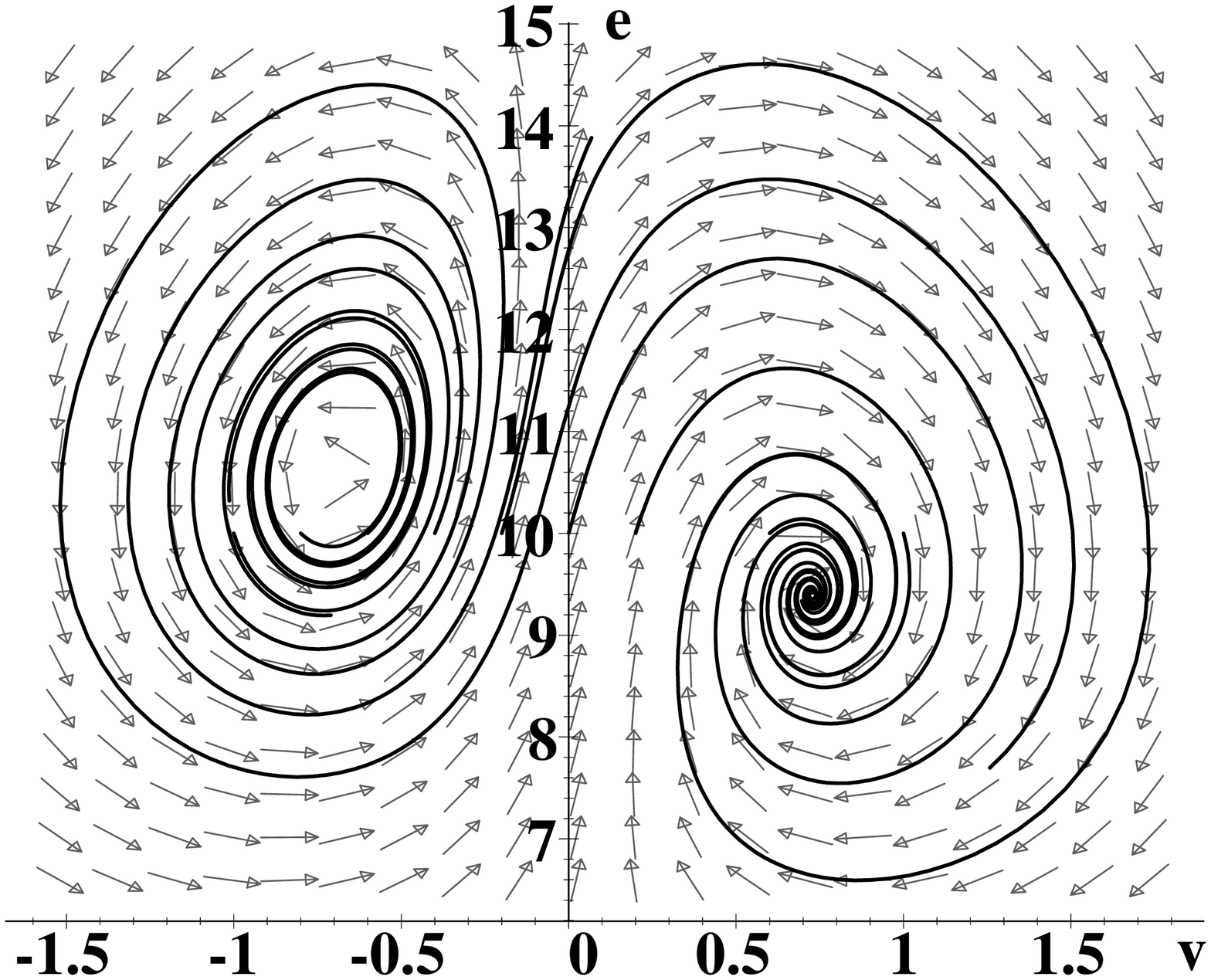,width=8cm,angle=-90}}
\vspace*{-1cm}
\centerline{\psfig{figure=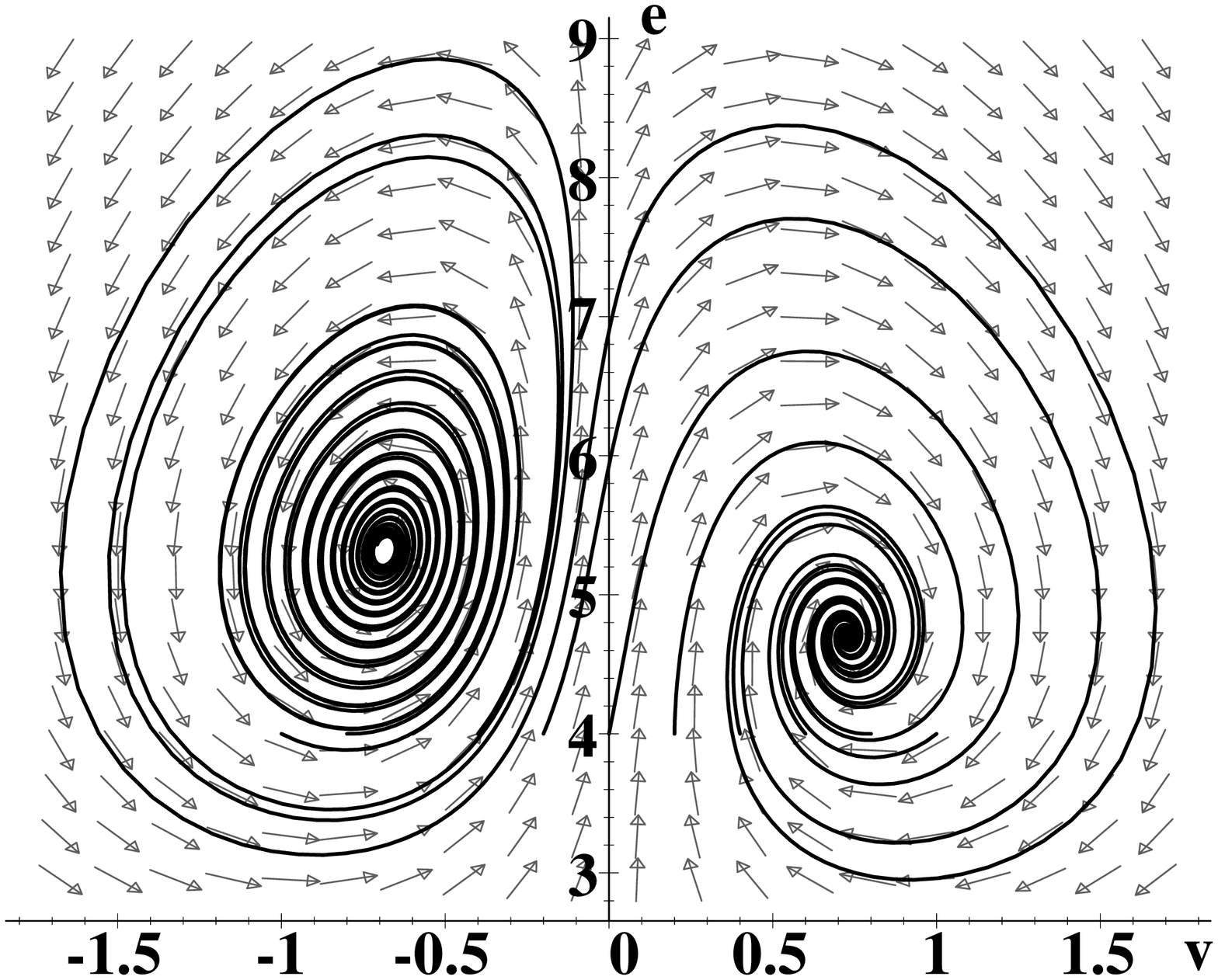,width=8cm,angle=-90}}
\caption[fig2]{
  Phase trajectories in the $\{v,e\}$ phase space for the motion into or
  against the direction of the driving force, which correspond either to
  positive or negative velocities. (a: top) $d_{2}=2.0$ corresponding to
  an instable ``uphill'' motion, (b: bottom) $d_{2}=4.0$ corresponding to
  a stable ``uphill'' motion, respectively.  Other parameters see
  \pic{bifurc}. \label{phase}}
\end{figure}

We find that below the bifurcation point which is
$d_{2}^{bif}=0.027$ for the given set of parameters, only one stable node
exists in the $\{v,e\}$ phase space, which corresponds to the passive
normal mode. Then, a subcritical bifurcation occurs which leads to 3
stationary solutions: a stable and an instable node, and a saddle point,
since all the $\lambda$ are real. At $d_{2}=0.046$, however, the nodes
turn into focal points.
With respect to the ``uphill motion'' we find in \pic{lambda12} 
the occurence of an
instable node at $d_{2}=0.027$, which then becomes an instable focus for $0.046 < d_{2}
< 2.720$. The respective real parts of $\lambda$ are equal in this range,
i.e. the $\lambda^{1,2}$ are complex.  The stability condition is
satisfied only if $\Re (\lambda) \leq 0$, which is above a second
critical value $d_{2}^{crit}=2.72$ for the given set of parameters.
That means, for $d_{2}>2.72$, the instable focal point becomes a stable
focus, which is also clearly shown in the phase plots of 
\pic{phase}a,b. In both Figures, we see a stable focal point for positive
values of the velocity, $v$, which correspond to the stable motion
``downhill'', i.e. in the direction of the driving force. For
$d_{2}=2.0<d_{2}^{crit}$, the phase plot for negative values of $v$ shows
an instable focal point, which turns into a stable focal point for
$d_{2}=4.0>d_{2}^{crit}$.

\begin{figure}[htbp]
  \centerline{\psfig{figure=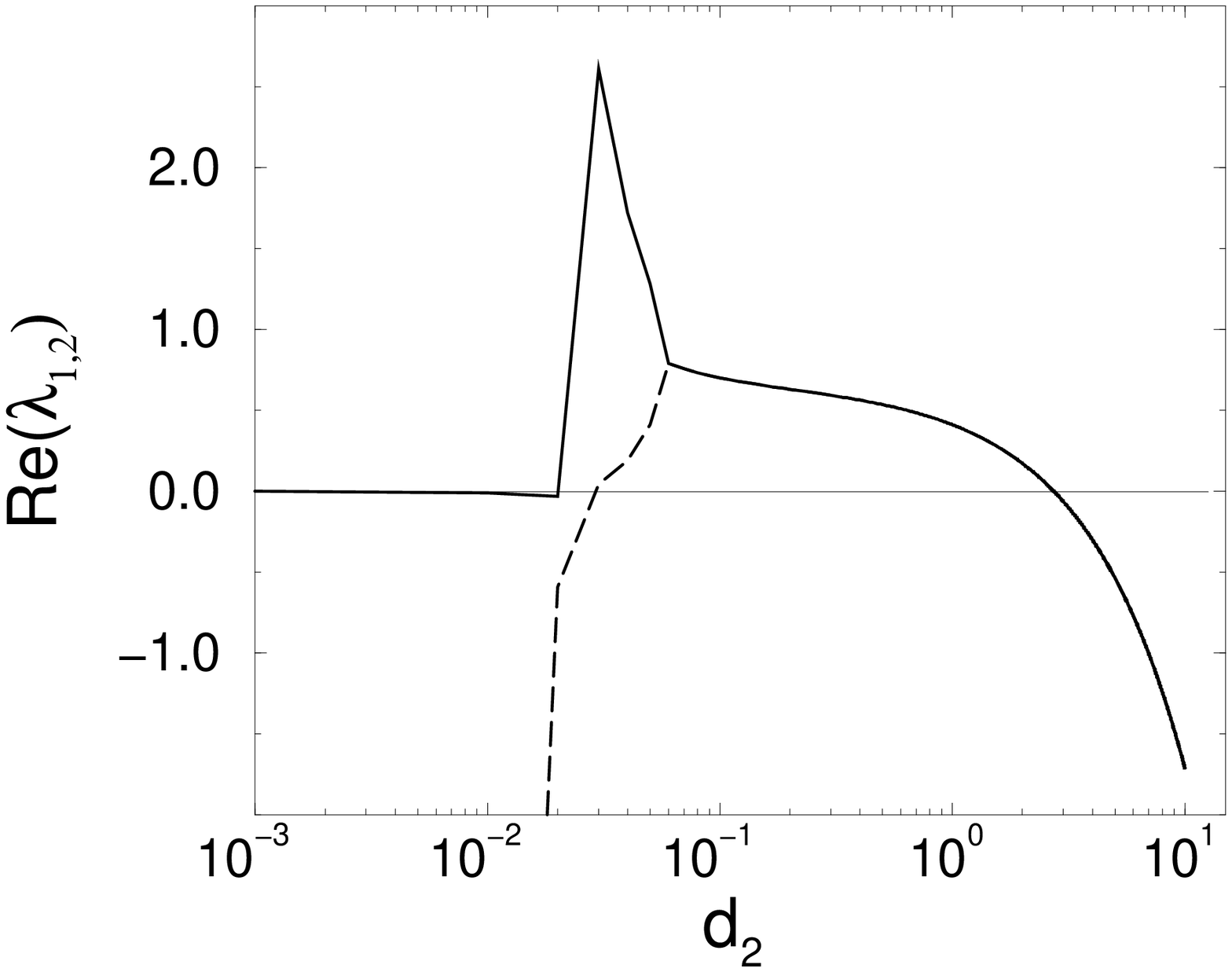,width=7.5cm}}
\caption[fig2]{
  Real part of $\lambda^{(1,2)}$, \eqn{lamb-12}, vs. conversion parameter
  $d_{2}$ for the stationary motion against the force, $\bbox{F}$.
  Parameters see \pic{bifurc}.  \label{lambda12}}
\end{figure}

Thus, we can conclude that for $d_{2}^{bif}<d_{2}$ an active mode of
motion becomes possible, which also implies the possibility of an
``uphill'' motion of the particle. However, only for values
$d_{2}^{crit}<d_{2}$, we can expect a \emph{stable} motion against the
direction of the force. 
The interesting result of a possible stable ``uphill-motion'' of
particles with internal energy depot will be employed in the following
section, where we turn to a more sophisticated, piecewise linear
potential.

For our further investigations, it will be useful to have a handy
expression for the critical supply of energy, $d_{2}^{crit}$, which
allows a stable ``uphill'' motion. This will be derived in the following,
with only a few approximations.  For the parameters used during the
computer simulations discussed later, \pic{lambda12} and Table I indicate
that the square root in \eqn{lamb-12} is imaginary, thus the stability of
the solutions depends on the condition:
\begin{equation}
  \label{c-stab}
  \gamma_{0}+ c+ d_{2}(v_{0}^{2}-e_{0}) \geq 0 
\end{equation}
If we insert the stationary value $e_{0}$, \eqn{v0-e0}, \eqn{c-stab}
leads to a 4th order inequation for $v_{0}$ to obtain stability:
\begin{equation}
  \label{v0-stab}
 (\gamma_{0} c-d_{2}q_{0}) \leq
v_{0}^{4}\;d_{2}^{2}+v_{0}^{2}\;(\gamma_{0} d_{2}+2cd_{2}) + c^{2}
\end{equation}
For a stable stationary motion of the particle, both \eqn{v0-cubic-2} and
\eqn{v0-stab} have to be satisfied.

The \emph{critical condition} for stability just results from the
equality in \eqn{v0-stab}, which then provides a replacement for the
prefactor $(\gamma_{0} c-d_{2}q_{0})$ in \eqn{v0-cubic-2}. If we insert
the critical condition into \eqn{v0-cubic-2}, we arrive at a 5th order
equation for $v_{0}$:
\begin{equation}
  \label{v0-5}
  \bbox{v}_{0}^{5}+\bbox{v}_{0}^{3}\,\left(\frac{2c}{d_{2}}\right)
+v_{0}^{2}\,\left(\frac{\bbox{F}}{d_{2}}\right)
+ \bbox{v}_{0}\,\left(\frac{c}{d_{2}}\right)^{2} 
+ \frac{c\bbox{F}}{d_{2}^{2}}=0 
\end{equation}
In order to simplify the further discussion, we assume that the internal
dissipation is negligible, $c=0$. Then, \eqn{v0-5} gives the simple
nontrivial solution:
\begin{equation}
  \label{v0-3-f}
  \bbox{v}_{0}^{3}=- \frac{\bbox{F}}{d_{2}} \;\;;\quad \mbox{if} \quad c=0
\end{equation}
This expression can be used to eliminate the stationary velocity,
$\bbox{v}_{0}$, in \eqn{v0-stab}. With the assumption $c=0$, we obtain
now from the critical condition, i.e. from the equality in \eqn{v0-stab},
a relation between the force, $\bbox{F}$, and the conversion parameter,
$d_{2}$.  Combining \eqn{v0-stab} and \eqn{v0-3-f} results in
\begin{equation}
  \label{d2-23}
  (-F)^{4/3}d_{2}\,+\,\gamma_{0}(-F)^{2/3}d_{2}^{2/3}\, -\,q_{0}
 d_{2}^{4/3}=0
\end{equation}
Because of $d_{2}>0$, the trivial and the negative solution of
\eqn{d2-23} can be neglected, and we finally arrive at the following
\emph{critical relation} for $d_{2}(F)$:
\begin{equation}
  \label{d2-f}
  d_{2}^{crit}=\frac{F^{4}}{8 q_{0}^{3}}\left(
1+\sqrt{1+\frac{4 \gamma_{0} q_{0}}{F^{2}}}\right)^{3}
\end{equation}
\pic{d-f} shows $d_{2}^{crit}$ as a function of the force, $F$.  In the
limit of negligible internal dissipation, the relation $d_{2}^{crit}(F)$
describes how much power has to be supplied by the internal energy depot
in order to allow a stable motion of the particle in \emph{both}
directions, in particular a \emph{stable uphill motion} of the particle.
\begin{figure}[htbp]
  \centerline{\psfig{figure=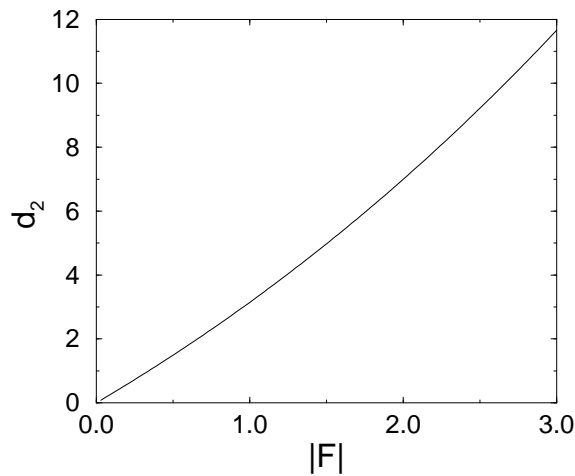,width=7.5cm}}
\caption[fig2]{
  Critical conversion rate, $d_{2}^{crit}$, \eqn{d2-f}, vs. amount of the
  driving force, $|F|$, to allow a stable motion of the particle both
  ``downhill'' and ``uphill'' (cf. \pic{flanke1}). For
  $d_{2}>d_{2}^{crit}$, the particle is able to move also against the
  direction of the force. Parameters see
  \pic{bifurc}.   \label{d-f}}
\end{figure}

\section{Deterministic Motion in a Ratchet Potential}
\label{4}
\subsection{Simulation Results for a Single Pumped Particle}
\label{4.1}

\begin{figure}[ht]
  \centerline{\psfig{figure=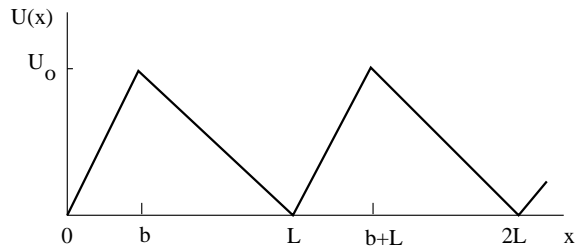,height=7.5cm,angle=-90}}
\caption[fig1]{
  Sketch of the asymmetric potential $U(x)$ (eq. (\ref{ux})). For the
  computer simulations, the following values are used: b=4, L=12, $U_{0}=
  7$ in arbitrary units.
\label{potent}}
\end{figure}

For further investigations of the motion of pumped particles, we specify
the potential $U(\bbox{x})$ as a piecewise linear, asymmetric potential
(cf.  Fig. \ref{potent}), which is known as a \emph{ratchet potential}:
\begin{eqnarray}
\label{ux}
U(x)=\left\{ 
\begin{array}{l}
\frac{\D U_0}{\D b}\{x-nL\} \\  $\quad$ \mbox{if $nL\leq x \leq nL+b$}\\ 
\frac{\D U_0}{\D L-b}\{(n+1)L-x\} \\  $\quad$ \mbox{if $nL+b \leq x \leq
  (n+1)L$}
\end{array}
\right. \\
(n=0,1,2,...) \nonumber
\end{eqnarray}

Further, we will use the following abbreviations with respect to the
potential $U(x)$, \eqn{ux}. The index $i=\{1,2\}$ refers to the two
pieces of the potential, $l_{1}=b$, $l_{2}=L-b$. The asymmetry parameter
$a$ should describe the ratio of the two pieces, and $\bbox{F}=
-\bbox{\nabla} U = const.$ is the force resulting from the gradient of
the piecewise linear potential. Hence, for the potential
$U(x)$, \eqn{ux}, the following relations yield:
\begin{eqnarray}
  \label{f-a}
 F_{1} &=& -\frac{U_{0}}{b}\;;\quad F_{2}=\frac{U_{0}}{L-b}\;;\quad 
a=\frac{l_{2}}{l_{1}}=\frac{L-b}{b}=-\frac{F_{1}}{F_{2}} \nonumber \\
F_{1}&=&-\frac{U_{0}}{L}(1+a)
\;;\quad F_{2} = \frac{U_{0}}{L}\frac{1+a}{a}
\end{eqnarray}
Whether or not the particle will be able to leave one of the potential
wells described by \eqn{ux}, depends in a first approximation on the
height of the potential barrier, $U_{0}$, and on the kinetic energy of
the particle. For particles with internal energy depot, the actual
velocity depends also on the conversion of internal into kinetic energy,
\eqn{model}. In agreement with the investigations in the previous
section, we can in principle distinguish between two different types of
motion: (i) a bound motion, i.e. the particle will not leave the
potential well because of the subcritical supply of energy from the
depot, but its position might oscillate within the boundaries, (ii) an
unbound motion, i.e. the particle will be able to leave the potential
well because of the supercritical supply of energy and move freely. Both
types of motion have analogies to the localized and delocalized states of
electrons in solid state physics as will be discussed in more detail in a
forthcoming paper.

In the following, we discuss computer simulations of the
\emph{deterministic} motion of {\em one} pumped Brownian particle in a
ratchet potential. The particle (mass $m=1$) starts its motion outside
the potential extrema; hence, there is an initial force on the particle.
The results for a single particle are shown in \pic{single-xve-t}, where
two different sets of parameters are used:
\begin{itemize}
\item[(i)] a small internal dissipation, $c$, which means a nearly ideal
  energy depot, and a large friction coefficient, $\gamma_{0}$, resulting
  in a \emph{strongly damped motion},
\item[(ii)] an internal dissipation, $c$, 10 times larger, an energy
  influx, $q_{0}$, ten times smaller, and a friction coefficient,
  $\gamma_{0}$, 100 times smaller than in (i), resulting in a \emph{weakly
    damped motion},
\end{itemize}

\begin{figure}[ht]
\centerline{\psfig{figure=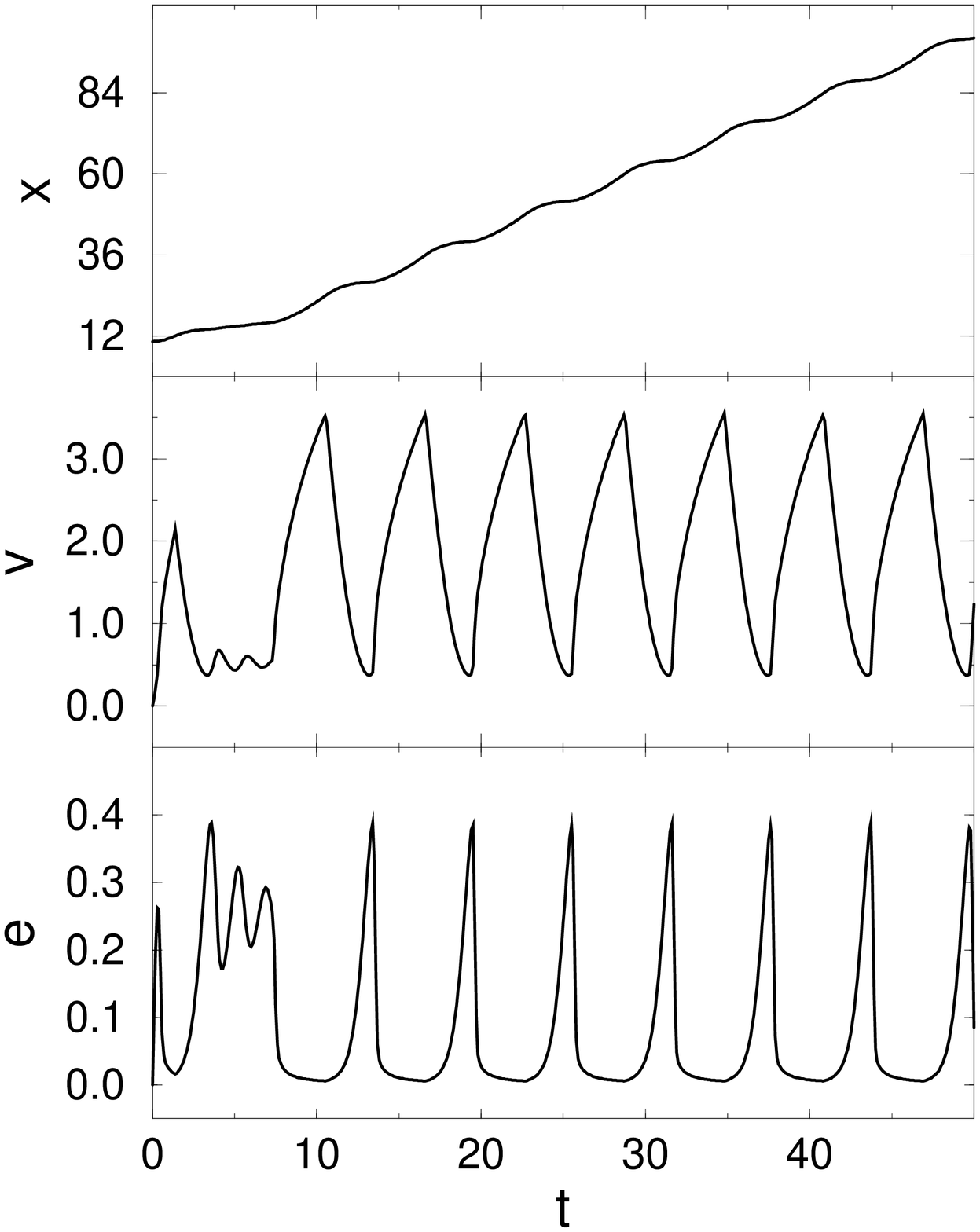,width=7.0cm}}
\vspace*{-1cm}
\centerline{\psfig{figure=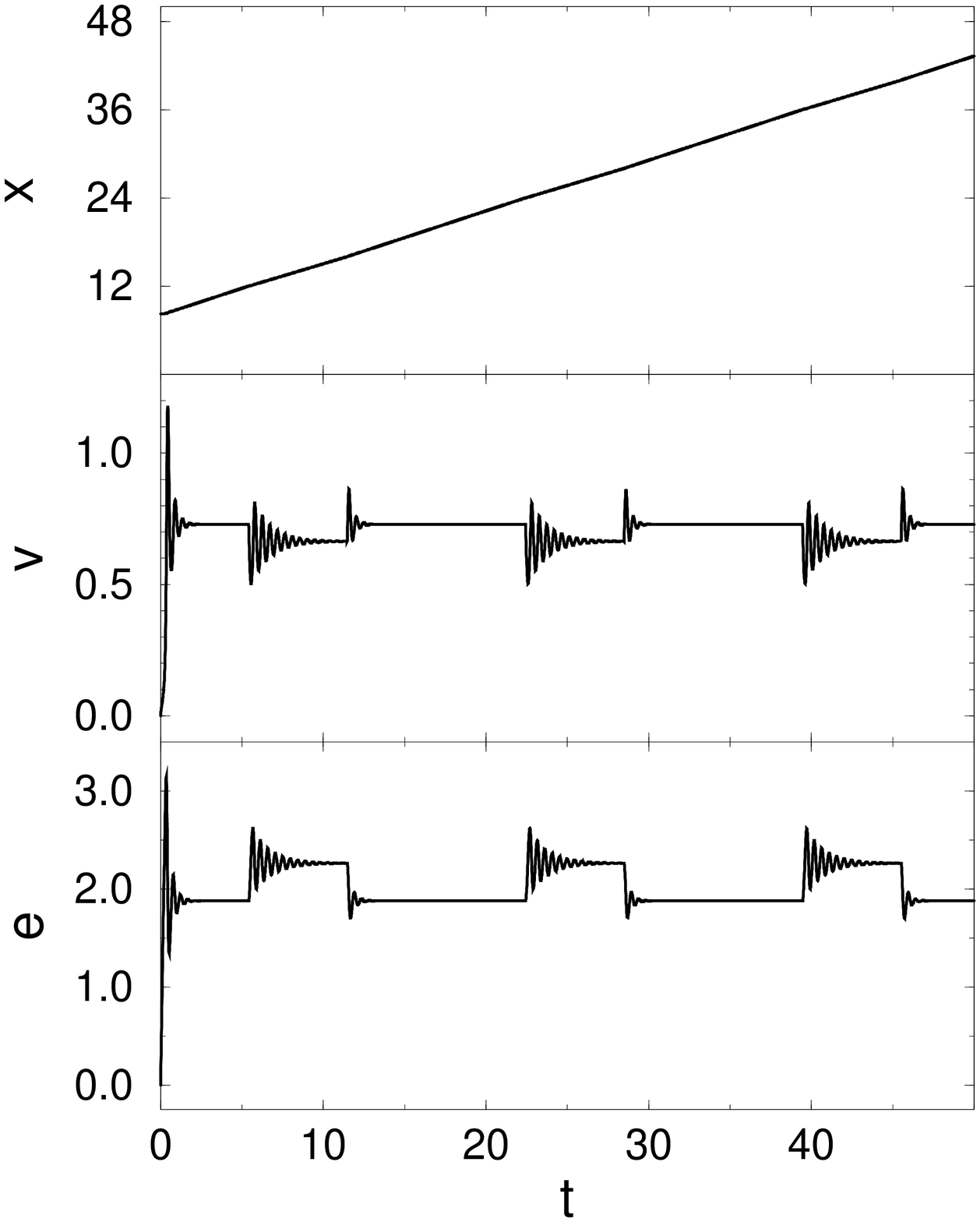,width=7.0cm}}
\vspace*{-1cm}
  \caption[fig2]{
    Trajectory $x(t)$, velocity $v(t)$ and energy depot $e(t)$ for a
    single particle moving in a ratchet potential (\pic{potent}).
    Parameters: (a: top) $q_0=1.0, \gamma = 0.2, c=0.1, d_2=14.0$,
    (b: bottom) $q_0=10, \gamma_{0}=20, c=0.01, d_2=10$. Initial
    conditions: $x(0) \in [4,12], v(0)=0, e(0)=0$}
 \label{single-xve-t}
\end{figure}

We note, that in the computer simulations always the complete set of
\eqn{model} for the particles is solved, regardless of the possible
approximations. The trajectories $\bbox{x}(t)$ in \pic{single-xve-t}
indicate a nearly uniform motion of the particle into one direction.
The continuous motion corresponds to a \emph{delocalized} state of the
particle in the ratchet potential, whereas for subcritical energetic
conditions only localized states exist.  The transition from localized to
delocalized states will be discussed in more detail in a forthcoming
paper \cite{tilch-fs-eb-99}.

As shown in \pic{single-xve-t}a, the less damped motion of a single
particle may result in steady oscillations in the velocity,
$\bbox{v}(t)$, and the energy depot, $e(t)$. Only if the damping is large
enough, the velocity and the energy depot may reach constant values (cf.
\pic{single-xve-t}b).  These values are of course different for each
piece of the potential, hence the periodical movement through every
maximum or minimum of the potential results in jumps both in the velocity
and the energy depot, which are followed by oscillations. In the phase
space shown in \pic{damp-traj}a, the motion of the particle appears as a
transition between two stable fix points, each of which describes the
stable motion on one flank.

\begin{figure}[ht]
  \centerline{\psfig{figure=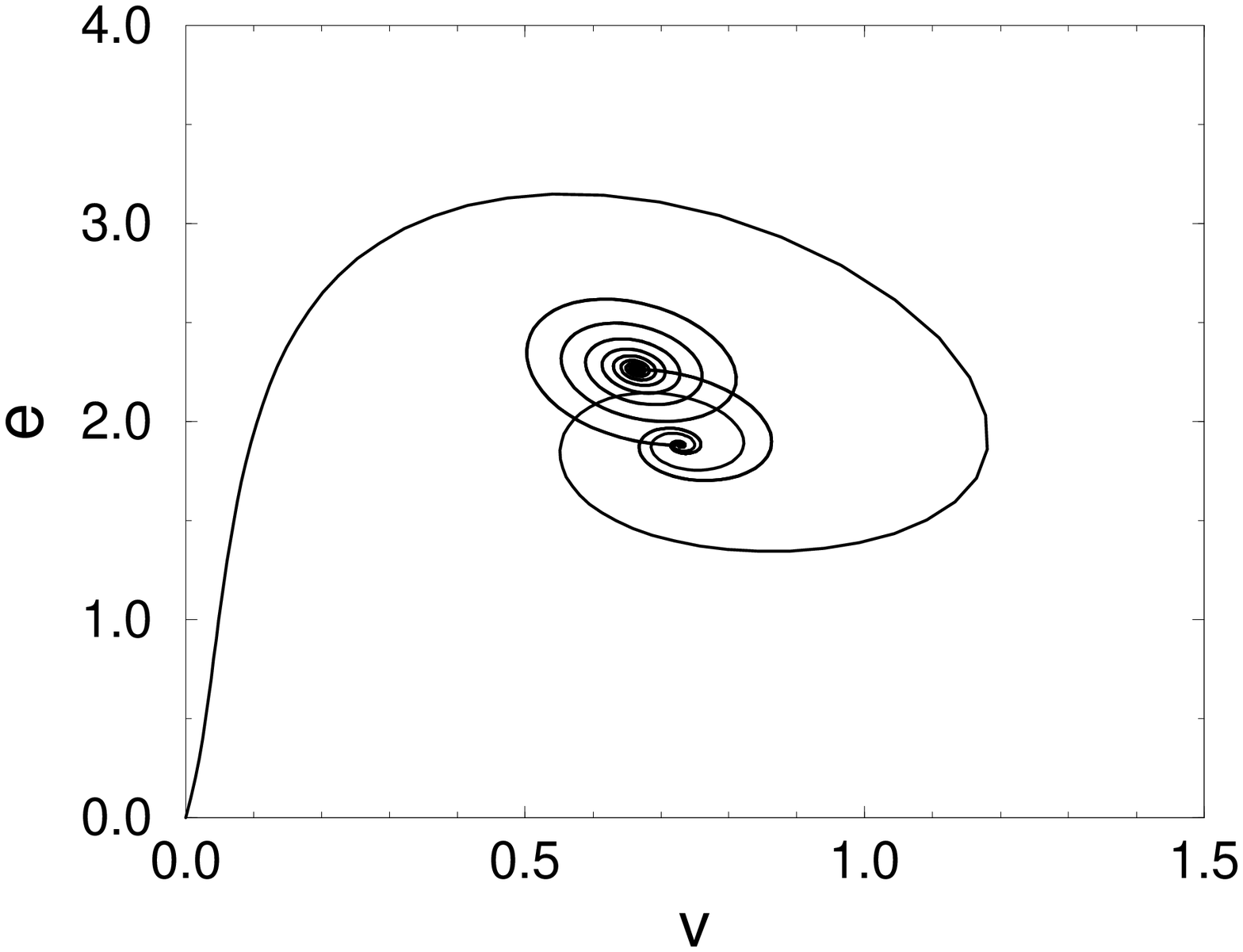,width=7.5cm}}

 \centerline{\psfig{figure=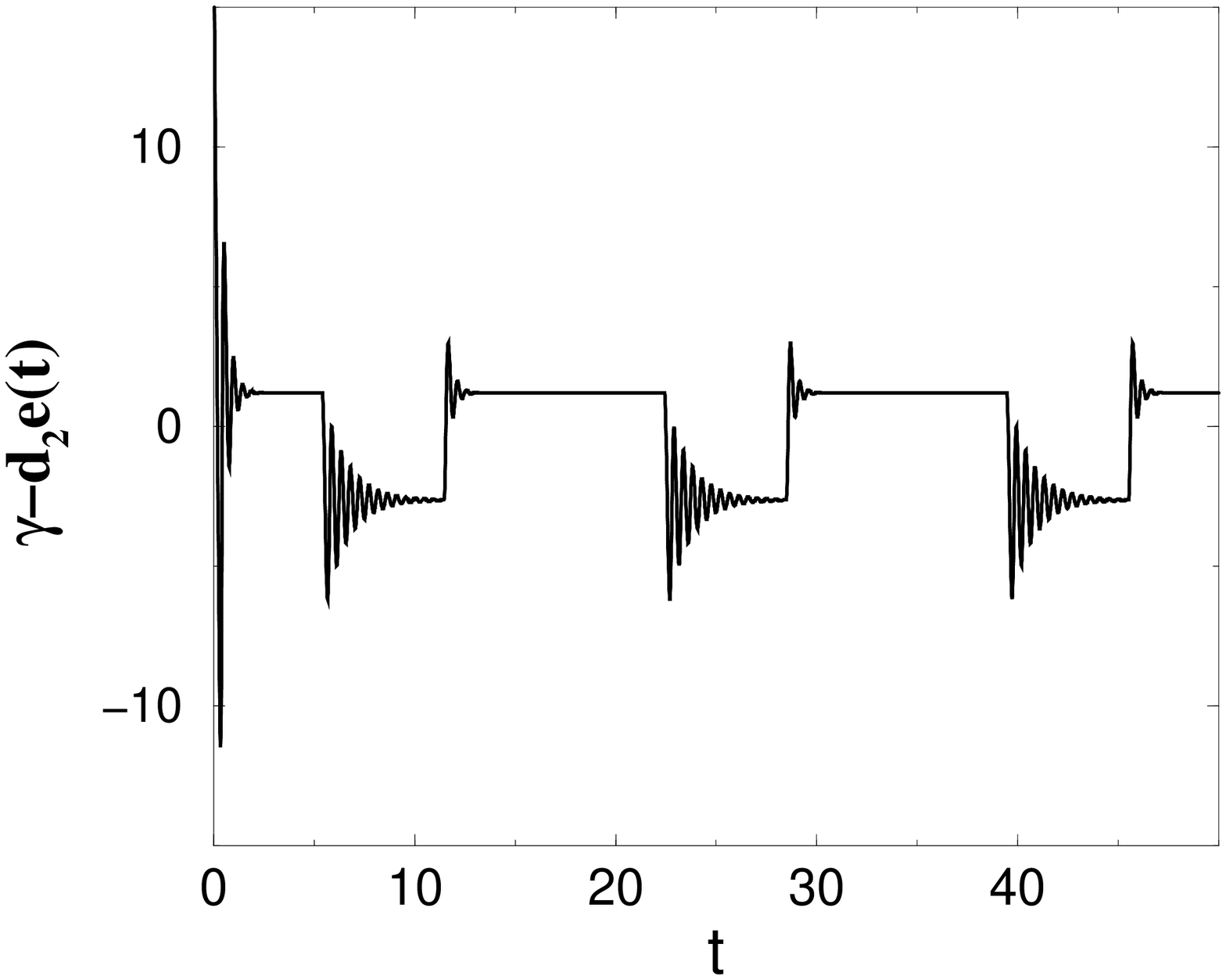,width=7.5cm}}
\caption[fig3]{
  (a: top) Energy depot $e(t)$ vs. velocity $v(t)$,  
  (b: bottom) prefactor: $\gamma - d_{2}e(t)$ for the strongly damped
  motion of a single particle (data from \pic{single-xve-t}b).
  \label{damp-traj}}
\end{figure}

In the following, we restrict the discussion to the \emph{strongly damped
  case}.  The oscillations which occur in $v$ and $e$ are damped out on a
characteristic time scale, $\tau= 1/\gamma_{0}$. If we assume that the
particle moves on the different pieces of the potential $\{b, \;L-b\}$
(cf.  \pic{potent}) during the two characteristic time intervals:
$T_{b}=b/v_{b}$ and $T_{L-b}= (L-b)/v_{(L-b)}$, then the particle is
subject to a constant force only as long as $\tau \ll T_{b}$ or $\tau \ll
T_{L-b}$, respectively.  For times larger than the characteristic time,
$\tau$, the motion of the particle can be described by the equation of
the overdamped limit, \eqn{x-overd}.  If we neglect again stochastic
influences, \eqn{x-overd} can be rewritten in the form:
\begin{equation}
  \label{v0-1}
 0 = -[\gamma_{0} - d_{2} e_{0}]\; \bbox{v}_0 + \bbox{F} 
\end{equation}
where $\abs{\dot{\bbox{x}}}=\abs{\bbox{v}_{0}}=const.$ is the velocity in
the overdamped limit and $\bbox{F}=\{F_{1},F_{2}\}$ is defined by
\eqn{f-a}. The stationary value for the internal energy depot, $e_{0}$,
is given by \eqn{v0-e0}.
If we assume again that the term $c\bbox{F}$ is small, then 
the constant velocity can be calculated from
\eqn{3stat-c0}, with $\bbox{F}$ specified as $F_{1}$ or $F_{2}$,
respectively. 
For any constant force $F$, there are two possible non-trivial solutions
of \eqn{3stat-c0}: a positive and a negative velocity with different, but
constant amount, which depend on the gradient of the potential.
The nontrivial values $v_{0}\neq 0$ of \eqn{3stat-c0} can be compared
with the constant values obtained in the simulations, and we find:
\begin{eqnarray}
  \label{veloc}
&\qquad \mbox{\bf \pic{single-xve-t}b} \qquad& \mbox{\bf \eqn{3stat-c0}} \non
\mbox{\bf lower value $v_{0}$} & 0.665 &\quad 0.6642 \\
\mbox{\bf upper value $v_{0}$} & 0.728 &\quad 0.7288 \no
\end{eqnarray}
The eqs. (\ref{x-overd}), (\ref{v0-1}) for the overdamped limit indicate
that the dynamics remarkably depends on the sign of the prefactor
$\gamma_{0} -d_2 e_{0}$, which governs the influence of the potential and
the stochastic force. Therefore, the prefactor should be discussed in
more detail, now.  \pic{damp-traj}b shows that the prefactor $\gamma_{0}
-d_2 e(t)$ displays a behavior similar to the velocity,
\pic{single-xve-t}b. The prefactor jumps between a positive and a
negative constant value, which can be approximated by means of the
constant, $e_{0}$, \eqn{v0-e0}, reached after the oscillations damped
out. It is shown that the jump occurs at the same time when the gradient
of the potential changes its sign. This can be also proved analytically.
Using \eqs{e0}{3stat-c0}, the prefactor $\gamma_{0}-d_2 e_{0}$ in
\eqn{x-overd} and \eqn{v0-1}, respectively, can be rewritten and we find
after a short calculation:
\begin{equation}
\label{pref-1}
\frac{1}{\gamma_{0} -d_2 e_{0}} = \frac{1}{2\gamma_{0} F_{i}} \left[F_{i}
\pm \sqrt{F_{i}^2+4\Big(q_0\gamma_{0}-c\gamma_{0}^{2}/d_{2}\Big)}\right] 
\end{equation}
This means that the product of the prefactor and the potential
gradient always has the same positive (or negative) sign, and the
direction of motion for the particle is only determined by the initial
condition. 

The prefactor $\gamma_{0} -d_2 e_{0}$ describes the balance between
dissipation and the energy supply from the internal depot of the particle.
Therefore it is expected that the time average of the prefactor 
should be zero, when averaged over one time period $T$: 
\begin{equation}
  \label{aver-2-a}
\mean{\gamma_{0} -d_2e_{0}} \,T = 0  
\end{equation}
This can be proven using $T=T_{b}+T_{L-b}$ as discussed above. The values
$v_{b}$, $v_{L-b}$ and $e_{b}$, $e_{L-b}$ for the related velocity and
energy depot along the different pieces of the potential can be
calculated by means of \eqs{3stat-c0}{v0-1} which leads directly to
\eqn{aver-2-a}.

\subsection{Investigation of the Net Current for an Ensemble of Particles}
\label{4.2}

Let us now discuss the \emph{deterministic} motion of an {\em ensemble}
of $N$ pumped Brownian particles in a ratchet potential. For the computer
simulations, we have assumed that the start locations of the particles
are equally distributed over the first period of the potential, $\{0,L\}$
and their initial velocity is zero.  In the deterministic case, the {\em
  direction of motion} and the {\em velocity} at any time $t$ are mainly
determined by the initial conditions. Hence, particles with an initial
position between $\{0,b\}$, which initially feel a force into the
negative direction, most likely move with a negative velocity, whereas
particles with an initial position between $\{b,L\}$ most likely move
into the positive direction.  This is also shown in \pic{damp-ve-x0}a,
where the velocity $v$ is plotted versus the initial position of the
particles.  Oscillations occur only at the minima and maxima of the
related potential, indicating a strong sensitivity to the initial
condition in these regions. The distribution of the final velocity is
shown in \pic{damp-ve-x0}b.

\begin{figure}[ht]
  \centerline{\psfig{figure=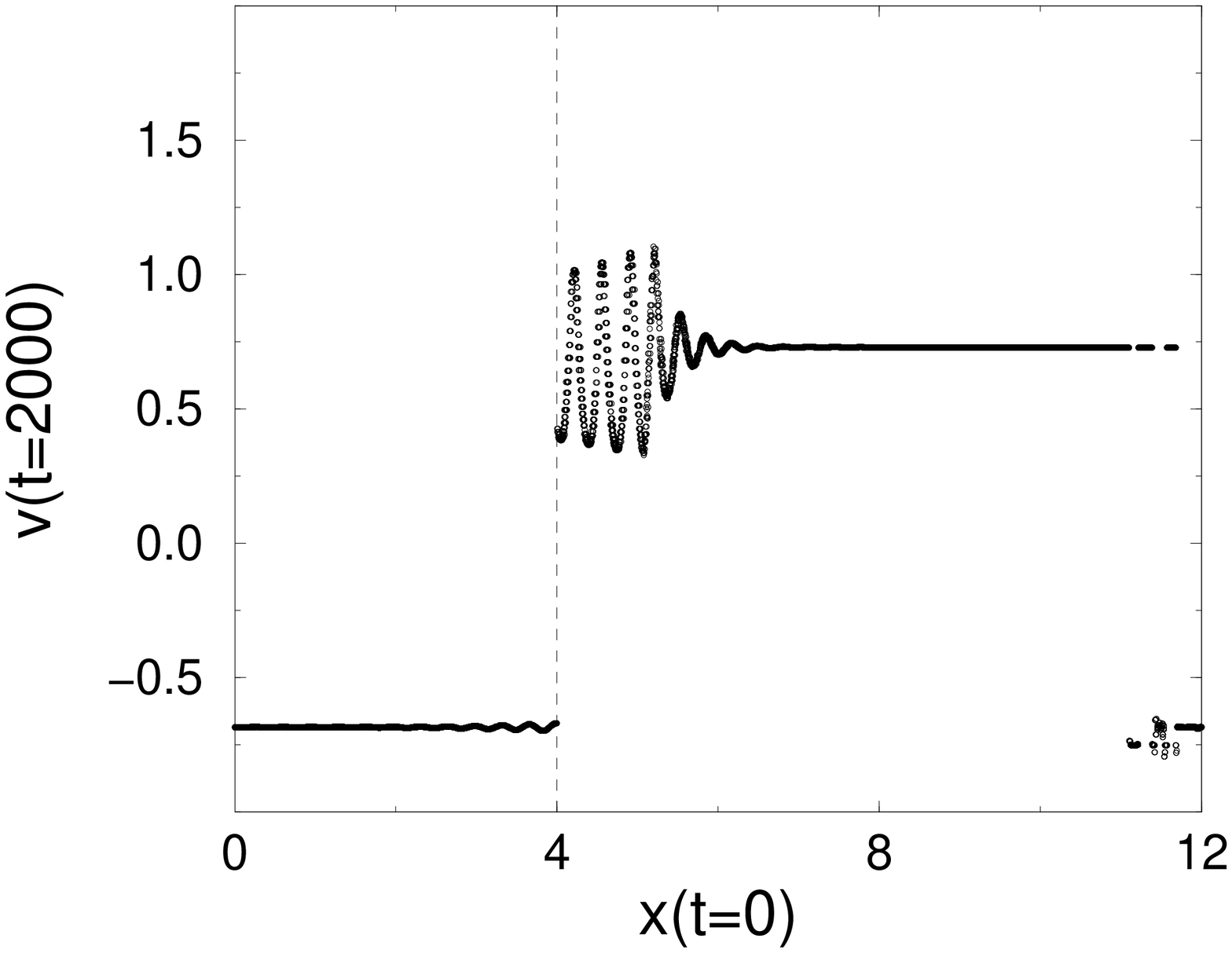,width=7.5cm}}

 \centerline{\psfig{figure=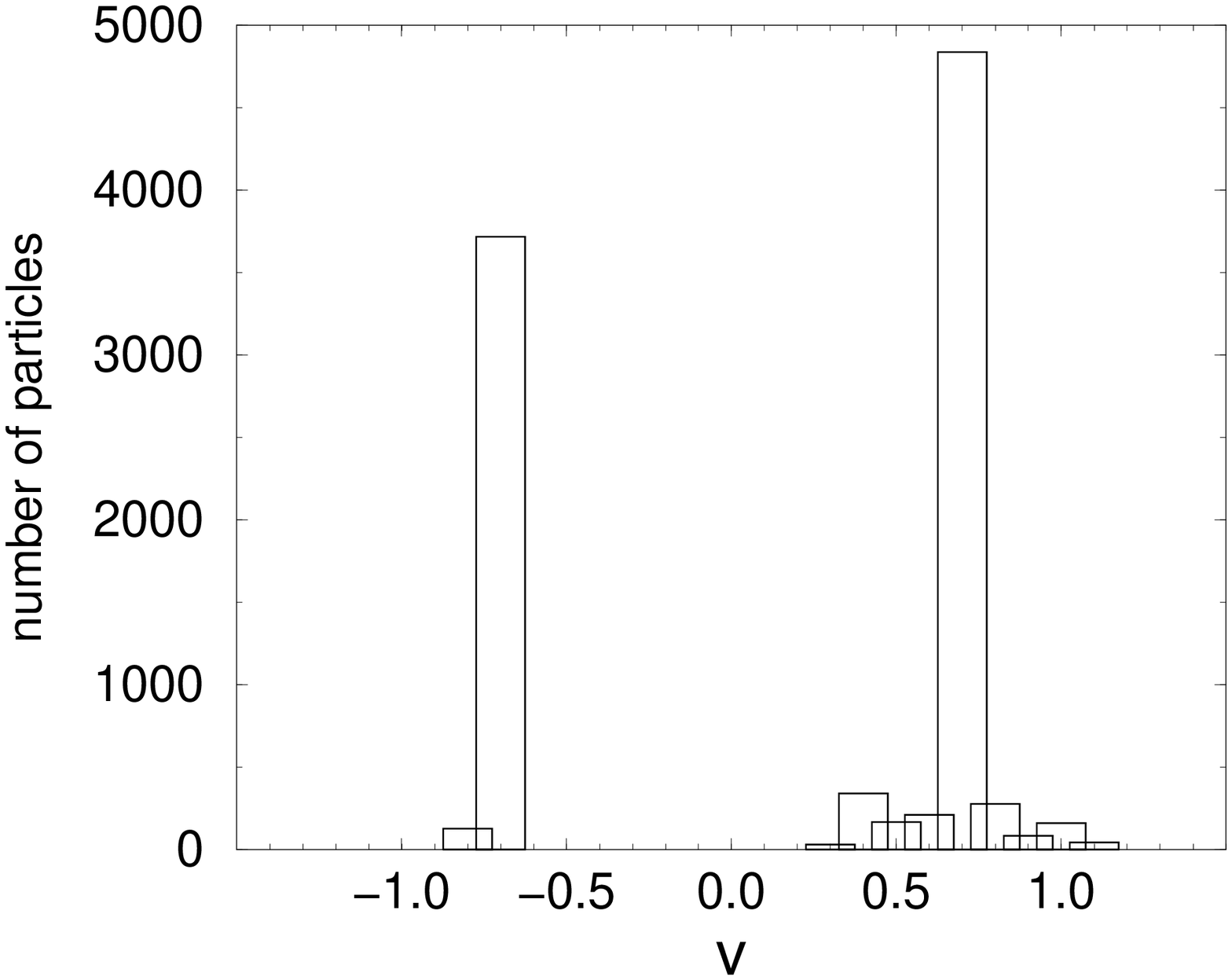,width=7.5cm}}
\caption[fig4]{
  (a: top) Final velocity $v_{e}$ after t=2.000 simulation steps
  (averaged for 10.000 particles) vs. initial location $x_{0}$ of the
  particles.  (b: bottom) Distribution of the final velocity $v_{e}$.
  Parameters see \pic{single-xve-t}b, initial locations of the particles
  are equally distributed over the first period of the ratchet potential
  $\{0,L\}$.
    \label{damp-ve-x0}}
\end{figure}

From \pic{damp-ve-x0}b we see two main currents of particles occuring,
one with a positive and one with a negative velocity, which can be
approximated by \eqn{3stat-c0}. The net current, however, has a positive
direction, since most of the particles start with the matching initial
condition. The time dependence of the averages is shown in
\pic{damp-avest}. The long-term oscillations in the average velocity and
the average energy depot result from the superposition of the velocities,
which are sharply peaked around the two dominating values
(cf. \pic{damp-ve-x0}b).

\begin{figure}[ht]
\centerline{
  \psfig{figure=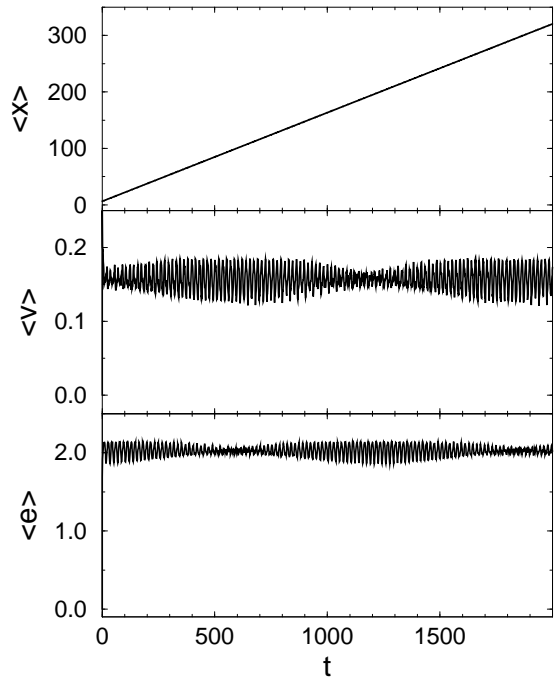,width=7.5cm}}
  \caption[fig5]{
    Averaged location $\langle x \rangle$, velocity $\langle v \rangle$
    and energy depot $\langle e \rangle$ of 10.000 particles vs. time
    $t$. Parameters see \pic{single-xve-t}b.}
  \label{damp-avest}
\end{figure}

The existence of periodic stationary solutions, $\bbox{v}_{0}(x) =
\bbox{v}_{0}(x \pm L)$, requires that the particles are able to escape
from the initial period of the potential. For a continuous motion which
corresponds to delocalized states, the particles must be able to move
``uphill'' on one or both flanks of the ratchet potential. In \sect{3},
we already investigated the necessary conditions for such a motion for a
single flank, and found a critical condition for the conversion rate,
$d_{2}$, \eqn{d2-f}. In order to demonstrate the applicability of
\eqn{d2-f} for the ratchet potential, we have investigated the dependence
of the \emph{net current}, expressed by the mean velocity $\mean{v}$, on
the conversion rate, $d_{2}$, for the overdamped case. The results of
computer simulations are shown in \pic{v-d2}.

\begin{figure}[htbp]
  \centerline{\psfig{figure=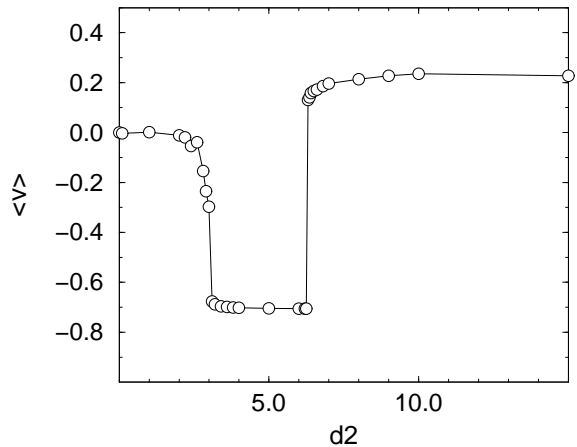,width=7.5cm}}
\caption{
  Average velocity $\langle v \rangle$ vs. conversion parameter $d_2$.
  The data points are obtained from simulations of 10.000 particles with
  arbitrary initial positions in the first period of the ratchet
  potential. Other parameters see \pic{single-xve-t}b. \label{v-d2}}
\end{figure}

In \pic{v-d2}, we see the existence of \emph{two different critical
  values} for the parameter $d_{2}$, which correspond to the onset of a
\emph{negative net current} at $d_{2}^{crit1}$ and a \emph{positive net
  current} at $d_{2}^{crit2}$.  For values of $d_2$ near zero and less
than $d_{2}^{crit1}$, there is no net current at all. This is due to the
subcritical supply of energy from the internal depot, which does not
allow an uphill motion on any flank of the potential. Consequently, after
the initial downhill motion, all particles come to rest in the minima of
the ratchet potential, with $v_{0}=0$ as the only stationary solution for
the velocity. With an increasing value of $d_{2}$, we see the occurence
of a negative net current at $d_{2}^{crit1}$. That means, the energy
depot provides enough energy for the uphill motion along the flank with
the lower slope, which, in our example, is the one with $F=7/8$ (cf.
\pic{potent}). If we insert this value for $F$ into the critical
condition, \eqn{d2-f}, a value $d_{2}^{crit1}=2.715$ is obtained, which
agrees with the onset of the negative current in the computer
simulations, \pic{v-d2}.

For $d_{2}^{crit1} \leq d_{2}\leq d_{2}^{crit2}$, a stable motion of the
particles up and down the flank with the lower slope is possible, but the
same does not necessarily apply for the steeper slope. Hence, particles
which start on the lower slope with a positive velocity, cannot continue
their motion into the positive direction since they are not able to climb
up the steeper slope. Consequently, they turn their direction on the
steeper slope, then move downhill driven by the force into the negative
direction, and continue to move into the negative direction while
climbing up the lower slope. Therefore, for values of the conversion rate
between $d_{2}^{crit1}$ and $d_{2}^{crit2}$, we only have an
\emph{unimodal} distribution of the velocity 
centered around the negative value:
\begin{equation}
  \label{v-neg}
v_1 = \frac{F}{2\gamma_{0}} - \sqrt{\frac{F^2}{4\gamma_{0}^2}
+\frac{q_0}{\gamma_{0}}} = -0.6855 \;\;\mbox{for}\;\; F=\frac{7}{8}
\end{equation}
which is independent of $d_{2}$ if the term $(c/d2)$ in \eqn{3stat-c0} is
negligible, which holds for the considered case.
For $d_{2}>d_{2}^{crit2}$, the energy depot also supplies enough energy
for the particles to climb up the steeper slope, consequently a periodic
motion of the particles into the positive direction becomes possible,
now. In our example, the steeper slope corresponds to the force $F=-7/4$
(cf. \pic{potent}) which yields a critical value $d_{2}^{crit1}=5.985$,
obtained by means of \eqn{d2-f}. This result agrees with the onset of the
positive current in the computer simulations, \pic{v-d2}.

For $d_{2}>d_{2}^{crit2}$, we have a \emph{bimodal} velocity
distribution, as also shown in \pic{damp-ve-x0}b. The net current,
which results from the average of the two main currents, has a positive
direction in the deterministic case, because most of the particles start
into a positive direction, as discussed above. We may simply assume,
that the number of particles in each direction is roughly proportional to
the length of the flank where they started from, which is also indicated
by the velocity distribution, \pic{damp-ve-x0}b. Then the mean
velocity in the strongly damped case can be approximated by:
\begin{equation}
  \label{mean-v}
  \mean{v}=\frac{1}{N} \sum_{i=1}^{N}v_{i}= \frac{1}{3}\,v_{1}
+\frac{2}{3}\,v_{2}
\end{equation}
where $v_{1}$ and $v_{2}$ are the stationary velocities on each flank,
which, in the limit of an nearly ideal energy depot, can be determined
from \eqn{3stat-c0}. With the negative velocity, $v_{1}$, \eqn{v-neg},
and the positive velocity, 
\begin{equation}
  \label{v-pos}
v_2 = \frac{F}{2\gamma_{0}} + \sqrt{\frac{F^2}{4\gamma_{0}^2}
+\frac{q_0}{\gamma_{0}}} = 0.664 \;\;\mbox{for}\;\; F=-\frac{7}{4}
\end{equation}
we find from \eqn{mean-v} for $d_{2}>d_{2}^{crit2}$ an average velocity,
$\mean{v}=0.216$, which also agrees with the computer simulations,
\pic{v-d2}. 
 
The results of the computer simulations have demonstrated that
in the deterministic case for the given special initial condition, i.e.
the equal distribution of particles over the ratchet period, the
direction of the net current can be adjusted by choosing the appropriate
values of the conversion rate, $d_{2}$.

The critical values for $d_{2}$, on the other hand, depend on the slope
of the two flanks of the potential, expressed by the force $F$. Lower
slopes also correspond to lower values of the conversion rate, because
less power is needed for the uphill motion.

We conclude our results by investigating the influence of the slope on
the establishment of a positive or negative net current. With a fixed
height of the potential barrier, $U_{0}$, and a fixed length $L$, the
ratio of the two different slopes is described by the asymmetry parameter
$a=l_{2}/l_{1}=-F_{1}/F_{2}$, \eqn{f-a}. The occurence of a current in
the ratchet potential requires the possibility of uphill-motion, which
depends on the critical supply of energy, described by \eqn{d2-f}.  In
order to obtain the critical value for the asymmetry of the potential, we
replace the force $F$ in \eqn{d2-f} by the parameter $a$, \eqn{f-a}. In
our example, the flank  $l_{1}$ of the potential has the steeper slope,
so the critical condition is determined by $F_{1}=U_{0}/L\;(1+a)$.  As
the result, we find:
\begin{equation}
  \label{a-crit1}
  a^{crit}=\frac{L}{U_{0}}\left[-\frac{\gamma_{0}}{2} d_{2}^{-1/3} +
\sqrt{\frac{\gamma^{2}}{4} d_{2}^{-2/3}+q_{0}d_{2}^{1/3}}
\;\right]^{3/2} -1 
\end{equation}
$a^{crit} \geq 1$ gives the critical value for the asymmetry, which may
result in a reversal of the net current. For $a>a^{crit}$, the flank
$l_{1}$ is too steep for the particles, therefore only a negative current
can occur which corresponds to the unimodal velocity distribution
discussed above. For $1<a<a^{crit}$, however, the particles are able to
move uphill either flank. Hence, also a positive current can establish
and the velocity distribution becomes bimodal, which results in a
positive net current.

\begin{figure}[htbp]
  \centerline{\psfig{figure=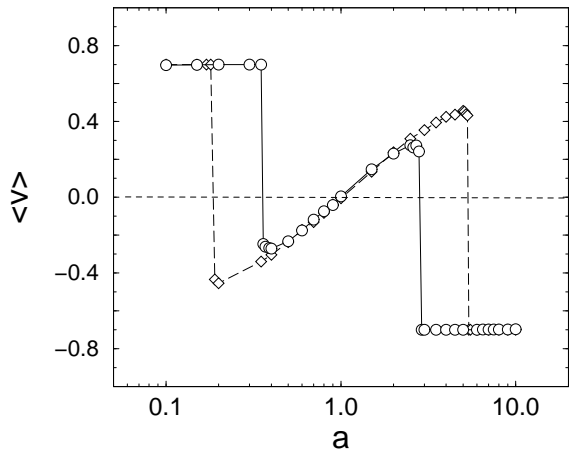,width=7.5cm}}
\caption{
  Average velocity $\mean{v}$ vs. asymmetry parameter $a$, \eqn{f-a}.
  The data points are obtained from simulations of 10.000 particles with
  arbitrary initial positions in the first period of the ratchet
  potential. $(\circ)$: $d_{2}=10$ $(-\!\!\!-\!\!\!-)$, $(\Diamond)$:
  $d_{2}=20$ $(--)$. Other parameters see \pic{single-xve-t}b.
  \label{v-as}}
\end{figure}

The current reversal from a negative to a positive net current is shown
in \pic{v-as}. Dependent on the value of the conversion rate, $d_{2}$, we
see the switch from the negative to the positive value of the net current
at a critical value of the asymmetry parameter $a$. Because of the
definition of $a$, the results for $a<1$, are the inverse of the results
for $a>1$. Obviously, for a symmetric ratchet, $a=1$ no net current
occurs, because the two main currents compensate. From \eqn{a-crit1}, we
obtain $a^{crit}=3.5$ for $d_{2}=10$, and $a^{crit}=6.5$ for $d_{2}=20$,
which both agree with the results of the computer simulations,
\pic{v-as}. Further, the results of \pic{v-as} shows that the stationary
velocities are independent of $d_{2}$ in the limit of an nearly ideal
energy depot, which is also indicated by \eqn{3stat-c0}.

\section{Summary and Conclusions}
\label{5}

In this paper, we have investigated the dynamics of Brownian particles
with an internal energy depot, which can be filled by the take-up of
energy from the environment. This extension which is inspired by the
biological features of active motion, has several aspects: 
(i) the
spatial or spatial-temporal inhomogeneous distribution of energy from the
environment can be considered, (ii) the supply of depot energy for
different activities, i.e. for accelerated motion or signal-response
behavior, can be modeled, (iii) in addition to external dissipative
processes caused by friction, internal dissipation can be included, (iv)
the adjustment of the time scale for the relaxation of the energy depot
allows to consider delay effects, i.e. for the acceleration of motion.

In this paper, we are mainly interested in the question how the internal
energy depot will affect the particle's motion in an external potential.
We found that provided some supercritical energetic conditions, the
Brown\-ian particles are able to move in a ``high velocity'' or active
mode of motion characterized by a velocity much larger than the
Stokes 
velocity. The latter one can be denoted as the \emph{passive} mode of
motion. In addition to stochastic forces, it is basically determined by
the response to the gradient of an external potential. One of the
possible active modes of motion can be understood as a continuation of
the passive or Stokes mode, the particles motion in the direction of the
force being accelerated.  

\emph{Additionally}, we also found an active mode which describes a
motion of the particle \emph{against} the gradient of an external
potential.  We have investigated the critical conditions for the
deterministic motion of the particle in the overdamped limit for the case
of a linear potential.  We found that the ``uphill'' motion is described
by two critical conversion rates. The bifurcation diagram shows that, at
a critical value $d_{2}^{bif}$, the possibility of an ``uphill'' motion
appears as a new solution for the stationary velocity, i.e.  ``uphill''
motion requires the existence of multiple stationary solution for the
velocity. However, this active mode remains unstable as long as the
conversion rate is below a second critical value $d_{2}^{crit}$,
\eqn{d2-f}, which depends on parameters describing the energy balance,
such as the friction coefficient, $\gamma_{0}$, and the influx of energy
into the depot, $q_{0}$, and on the gradient of the potential, i.e. the
resulting force $\bbox{F}$. A \emph{stable} ``uphill'' motion occurs only
for a supercritical conversion rate.

These results allow us to interpret the deterministic motion of an
ensemble of particles in a ratchet potential, i.e. a piecewise linear,
periodic asymmetric potential with two different slopes. Initially, the
particles are equally distributed over the first period of the potential.
In order to produce a current, the particles must be able to escape from
the potential well, which depends on the supply of energy. We are able to
find two different critical conversion rates, which describe the onset of
a directed current into the different directions. As long as the
particles are only able to move uphill the flank with the lower slope, we
have an unimodal velocity distribution, corresponding to a negative net
current.  But if the particles are able to move uphill either flank, we
find a bimodal velocity distribution, and a positive net current for the
given conditions. Hence, in the deterministic case the
direction of the net current can be controlled via a single parameter,
$d_2$, which describes the conversion of internal into kinetic energy.
We are able to calculate the critical parameter, as
well as the resulting velocity of the net current from analytical
considerations, which agree with the results found in the computer
simulations. Alternatively, the critical supply of energy can be also
reformulated in terms of a critical asymmetry of the ratchet potential,
which restricts the occurence of a (positive or negative) net current.

To conclude the results obtained, we have shown that an ensemble of
pumped Brownian particles moving in a ratchet potential can produce a
\emph{directed net current}.  In this respect, our result agrees with the
conclusions of other physical ratchet models which have been proposed to
reveal the microscopic mechanisms resulting in directed movement.  Due to
\cite[p.  295]{haenggi-bart-96} a \emph{ratchet system} which is meant to
be ``a system that is able to transport particles in a periodic structure
with nonzero macroscopic velocity although on \emph{average} no
macroscopic force is acting.'' Indeed, we have shown in \eqn{aver-2-a}
that for the stationary approximation $\mean{\gamma_{0} -d_2e_{0}} \,\tau
= 0$ holds, i.e. the force acting on the particle is of zero average with
respect to one period, $\tau$. However, different from other ratchet
models, ie. the rocking ratchet
\cite{Bartussek-Haenggi-Kissner-94,Lindner-et-97,landa-98} or the
diffusion ratchet \cite{Reimann-et-96} which assume a spatially uniform
time-periodic force, the force in our model switches between two constant
values, dependent on the moving direction and the flank the particle is
moving on. Hence, it does not represent a spatially uniform force, and we
may conclude that the mechanism of motion which \eqn{x-overd} is based
on, should be different from the previous mechanisms which originate
directed motion in a ratchet potential.

In this paper, we have restricted the discussion to the deterministic
motion of Brownian particles with an internal energy depot. We note that
the stochastic motion of an ensemble of Brownian particles in a ratchet
potential will be investigated in detail in a forthcoming paper
\cite{tilch-fs-eb-99}, which pays particular attention to the influence
of stochastic effects on the establishment of the net current.
Additionally, we will also discuss the non-equilibrium distribution
functions for Brownian particles with internal energy depot
\cite{ebeling-et-99}.

As a final remark, we note in a more general sense that the introduction
of an internal energy depot adds an interesting and new element to the
known model of Brownian particles, which can be useful in two different
respects. With respect to the physical aspects, the dynamical system now
has a \emph{new degree of freedom}, which increases the phase space,
$\Gamma=\{x_{i},v_{i},e\}$.  Due to the additional energy supply, the
system is driven into non-equilibrium.
Hence, a qualitative new behavior in the particle's motion can be
obtained.

With respect to possible biological aspects, the extension of the
Brownian particle model by mechanisms of energy take-up, storage and
conversion, should contribute to the development of a \emph{microscopic
  theory} of active biological motion. The final goal of such a project
could be a microscopic image of well known phenomenological models of
biological motion, taking into account energy balances that are related
to the mechanisms of energy pumping and energy dissipation.


\end{document}